\documentclass[sn-nature,iicol]{sn-jnl}% Default with double column layout

\usepackage{graphicx}%
\usepackage{multirow}%
\usepackage{amsmath,amssymb,amsfonts}%
\usepackage{amsthm}%
\usepackage{mathrsfs}%
\usepackage[title]{appendix}%
\usepackage[table]{xcolor}
\usepackage{textcomp}%
\usepackage{manyfoot}%
\usepackage{booktabs}%
\usepackage{algorithm}%
\usepackage{algorithmicx}%
\usepackage{algpseudocode}%
\usepackage{bm}
\usepackage{makecell}
\usepackage{listings}%
\let\ten\bm
\DeclareMathOperator{\tr}{tr}

\begin{document}
%%%%%%%%%%%%%%%%%%%%%%%%%%%%%%%%%%%%%%%%%%%%%%%%%%%%%%%%%%%%%%%%%%%%%%%%%%%%
\newcommand{\beq}{\begin{equation}}
\newcommand{\eeq}{\end{equation}}
\newcommand{\turbo}{{{\textsf{FF\,TURBO$^{\sf{TM}}$\,PLUS}\,}}}
\newcommand{\leap}{{{\textsf{FF\,LEAP$^{\sf{TM}}$}\,}}}
\newcommand{\edge}{Edge\,}
\newcommand{\sky}{Sky\,}
\newcommand{\ray}{Ray\,}
%\newcommand{\metaspeededge}{{\textsf{METASPEED$^{\sf{TM}}$}\,EDGE}\,}
%\newcommand{\metaspeedsky}{{\textsf{METASPEED$^{\sf{TM}}$}\,SKY}\,}
%%%%%%%%%%%%%%%%%%%%%%%%%%%%%%%%%%%%%%%%%%%%%%%%%%%%%%%%%%%%%%%%%%%%%%%%%%%%
%\title[Mechanics of Super Shoes]{Discovering Mechanical Properties of Super Shoes}
\title[Discovering the mechanics of ultra-low density
elastomeric foams]{\sffamily{\bfseries{\hspace*{0.5cm} Discovering the mechanics of ultra-low density \\
elastomeric foams in elite-level racing shoes}}}
%%%%%%%%%%%%%%%%%%%%%%%%%%%%%%%%%%%%%%%%%%%%%%%%%%%%%%%%%%%%%%%%%%%%%%%%%%%%
\author*[]{\fnm{Jeremy A.} \sur{McCulloch}}\email{jmcc@stanford.edu}
\author[]{\fnm{Scott L.} \sur{Delp}}\email{delp@stanford.edu}
\author[]{\fnm{Ellen} \sur{Kuhl}}\email{ekuhl@stanford.edu}
\affil[]{\orgdiv{Departments of Mechanical Engineering and Bioengineering}, \orgname{Stanford University}, \orgaddress{\street{318~Campus Drive}, \city{Stanford}, \postcode{94305}, \state{CA}, \country{USA}}}
%150 to 250 words
\abstract{Ultra-low-density elastomeric foams 
enable lightweight systems 
that combine high compliance with efficient energy return.
Their mechanical response 
is inherently complex, 
characterized by 
high compressibility, 
nonlinear elasticity, and 
microstructural heterogeneity.
%They display a complex mechanical behavior 
%including
%high\,compressibility, 
%nonlinear\,elasticity, and 
%microstructural\,heterogeneity. 
%
In high-performance racing shoes, 
these foams are critical for 
low weight, 
high cushioning, and 
efficient energy return; 
yet, their constitutive behavior 
remains difficult to model and poorly understood. 
Here we integrate 
mechanical testing and 
machine learning
to discover the mechanics 
of two ultra-low density elastomeric polymeric foams
used in elite-level racing shoes. 
Across 
uniaxial tension, 
confined and unconfined compression, and 
simple shear, 
both foams 
exhibit pronounced tension–compression asymmetry, 
negligible lateral strains 
consistent with an effective Poisson's ratio close to zero, and 
low hysteresis indicative of an efficient energy return. 
Quantitatively, 
both foams provide a similar compressive stiffness
(268$\pm$16\,kPa vs. 299$\pm$29\,kPa),
while one foam exhibits a 42\% higher tensile stiffness
(884$\pm$69\,kPa vs. 623$\pm$96\,kPa), 
and nearly double the shear stiffness
(219$\pm$20\,kPa vs. 117$\pm$24\,kPa), 
implying a substantially greater lateral stability
at a comparable vertical energy return 
(83.3$\pm$1.5\% vs. 88.9$\pm$1.8\%). 
By integrating these data into 
constitutive neural networks, 
paired with sparse regression, 
we discover compact, interpretable 
single-invariant models--supplemented 
by mixed-invariant or principal-stretch based terms--that 
capture the unique signature of the foams with 
${\textsf{R}}^{\textsf{2}}$ values 
close to one across all loading modes. 
From a human performance perspective, 
these models enable finite-element and gait-level simulations 
of high-performance racing shoes 
to quantify 
running economy, 
performance enhancements, and 
injury risks on an individual athlete level. 
More broadly, 
this work establishes 
a scalable and interpretable approach 
for constitutive modeling of 
highly compressible, 
ultra-light elastomeric foams
with applications to
wearable technologies, 
soft robotics, and 
energy-efficient mobility systems.}
\maketitle
%%%%%%%%%%%%%%%%%%%%%%%%%%%%%%%%%%%%%%%%%%%%%%%%%%%%%%%%%%%%%%%%%%%%%%%%%%%%
\section{{\sffamily{\bfseries{Introduction}}}}
\label{sec:background}
%%%%%%%%%%%%%%%%%%%%%%%%%%%%%%%%%%%%%%%%%%%%%%%%%%%%%%%%%%%%%%%%%%%%%%%%%%%%
%%%%%%%%%%%%%%%%%%%%%%%%%%%%%%%%%%%%%%%%%%%%%%%%%%%%%%%%%%%%%%%%%%%%%%%%%%%%
%{\sffamily{\bfseries{Next-generation\,high-performance\,racing\,shoes.}}}
%\subsection{Carbon fiber plated running shoes}
%%%%%%%%%%%%%%%%%%%%%%%%%%%%%%%%%%%%%%%%%%%%%%%%%%%%%%%%%%%%%%%%%%%%%%%%%%%%
The two-hour marathon barrier 
remains a defining frontier 
of human performance.
The current world record of 2:00:35 
was set at the 2023 Chicago Marathon by Kelvin Kiptum 
%was set on October 8, 2023 by Kelvin Kiptum 
wearing a prototype of the Nike Alphafly 3, 
a high-performance racing shoe 
with a carbon-fiber plate 
embedded in an ultra-low density elastomeric foam 
\cite{fernandes_kiptum_2024}.
Over the past decade, 
the introduction of racing shoes 
that combine carbon-fiber plates and ultra-light elastomeric foams 
has been associated with substantial performance improvements 
among both recreational and elite athletes, 
particularly in long-distance running
\cite{herzog_secrets_2022}. 
Compared to earlier racing shoes, 
carbon-fiber-plated footwear exhibits a
reduced weight, 
increased cushioning, and 
high energy return,
%elevated longitudinal bending stiffness, 
features that collectively improve running economy 
by altering lower-limb mechanics 
and reducing energy dissipation 
\cite{hoogkamer_comparison_2018}. \\[6.pt]
% Consist of rubber outer sole, compliant foam, carbon fiber plate, compliant foam, insole, light mesh upper
%%%%%%%%%%%%%%%%%%%%%%%%%%%%%%%%%%%%%%%%%%%%%%%%%%%%%%%%%%%%%%%%%%%%%%%%%%%%
{\sffamily{\bfseries{Increasing the performance of next-generation racing shoes.}}}
%%%%%%%%%%%%%%%%%%%%%%%%%%%%%%%%%%%%%%%%%%%%%%%%%%%%%%%%%%%%%%%%%%%%%%%%%%%%
Extensive studies of the first prominent carbon-fiber-plated shoe,
the Nike Vapor-fly,
suggest that it can 
reduce the metabolic cost of running 
at a fixed speed by up to 4\% \cite{whiting_metabolic_2022}, 
or, equivalently, 
increase the speed by 4\% 
at constant metabolic demand \cite{patoz_vaporfly_2022}.  
Inspired by these early success stories, 
various brands are now producing 
their own versions of carbon-fiber-plated shoes 
in an attempt to 
increase speed and
decrease metabolic cost \cite{munizpardos_technological_2021}.
A recent comparison 
between seven carbon-fiber-plated shoes 
and one conventional shoe 
revealed that 
the Nike Vaporfly~2, 
the Nike Alphafly, and 
the Asics Metaspeed Sky 
achieved the greatest reduction 
in metabolic demand \cite{joubert_comparison_2021}. 
The study concluded 
that the longitudinal bending stiffness 
played a minor role in metabolic savings, and that,
with all shoe weights nearly equal, 
the greatest differentiating factor  
was the midsole foam. 
Increasing evidence suggests 
that the midsole foam--rather than the carbon-fiber plate--may hold the key 
to breaking the two-hour marathon barrier
\cite{rodrigocarranza_midsole_2024}.\\[6.pt]
%In particular, 
%it is important that the foam effectively store energy 
%and return it to the runner during toe off. 
%%
%\textcolor{red}{replace this by ``this raises the question... what's more important the carbon fiber plate or the foam?'' 
%A more recent study 
%compared two versions of the Asics Metaspeed Sky, 
%which primarily differ in the shape and position 
%of the carbon fiber plate in the shoe \cite{xu_effects_2025}. 
%%
%While this study did not explicitly measure the metabolic cost of running, 
%it showed that different plate positions 
%can alter the biomechanics of running. 
%%
%In particular, 
%the curved carbon fiber plate 
%in the Metaspeed Sky Tokyo shoe
%resulted in reduced 
%joint range of motion and 
%joint moment in the hip and knee, 
%suggesting that this newer model 
%may result in more efficient running.}  
%%%%%%%%%%%%%%%%%%%%%%%%%%%%%%%%%%%%%%%%%%%%%%%%%%%%%%%%%%%%%%%%%%%%%%%%%%%%
{\sffamily{\bfseries{Reducing injury risk by balancing cushioning and stability.}}}
%%%%%%%%%%%%%%%%%%%%%%%%%%%%%%%%%%%%%%%%%%%%%%%%%%%%%%%%%%%%%%%%%%%%%%%%%%%%
While the major focus of carbon-fiber-plated racing shoes 
lies in increasing running performance, 
it is equally important to understand 
whether these shoes might increase  
the prevalence of running-related injuries. 
Some studies have proposed 
that more energy is stored in the shoe 
and less is stored in muscles and tendons, 
which may decrease the risk of running injuries \cite{hoogkamer_comparison_2018}.
However, a recent case study of six stress fractures--sustained while wearing carbon-fiber-plated shoes--suggests that these shoes 
increase displacements of the navicular and cuneiform bones and 
modify forces to the hindfoot, 
which could increase injury risk\cite{tenforde_bone_2023}. 
By design, carbon-fiber–plated racing shoes 
are primarily optimized for straight-line running. 
The ultra-low density foams of their midsole 
exhibit an inherently low shear stiffness 
as a consequence of cellular solids scaling, 
meaning the shear modulus decreases 
super-linearly with decreasing density \cite{gibson_1997}.
This low shear stiffness  
reduces stability in cornering and uneven terrain, 
and increases injury risk.
%
%While this study did not test carbon-plated shoes, 
%an earlier study found counterintuitively 
%that the ground reaction forces were greater 
%in highly cushioned running shoes 
%when compared to conventional running shoes \cite{kulmala_running_2018}. 
%
%In the design and analysis 
%of carbon plated running shoes and foams, 
%it is important to consider 
%not only the impact on performance 
%but also the effect on injury risk. 
%%%%%%%%%%%%%%%%%%%%%%%%%%%%%%%%%%%%%%%%%%%%%%%%%%%%%%%%%%%%%%%%%%%%%%%%%%%%
\\[6.pt]
%%%%%%%%%%%%%%%%%%%%%%%%%%%%%%%%%%%%%%%%%%%%%%%%%%%%%%%%%%%%%%%%%%%%%%%%%%%%
{\sffamily{\bfseries{%Computational modeling of gait, shoes, foams.
Constitutive accuracy as the missing link in gait simulation.}}}
%%%%%%%%%%%%%%%%%%%%%%%%%%%%%%%%%%%%%%%%%%%%%%%%%%%%%%%%%%%%%%%%%%%%%%%%%%%%
Computational models of gait 
provide valuable guidance how 
to increase performance or reduce injury risk. 
The ground reaction forces and the joint angles 
of the hip, knee, and ankle are straightforward to measure, 
and can be integrated into 
a generative machine learning model \cite{tan_gaitdynamics_2025}; 
however, this model and many others 
neglect the mechanics of the foot and the shoe. 
Current attempts to model foot and shoe 
take magnetic resonance images \cite{zollner_highheels_2015}, 
segment the image stack, 
assign constitutive models to the different materials, and 
impose boundary conditions \cite{song_systematic_2025}. 
In this process, 
the major source of error is 
%the choice of boundary conditions and 
the selection of constitutive models for the foot and the shoe. 
This challenge is reflected 
in a study, which models the foot and the shoe 
using a six-degree of freedom model: 
The model accurately reconstructs joint kinematics, 
but predicts inaccurate ground reaction forces, 
and its results vary greatly 
for small variations in running terrain \cite{hannah_evaluation_2016}. 
Thus, further improvements in both
biomechanical modeling of the foot \cite{yong_foot_strike_2020} and 
constitutive modeling of the shoe \cite{yang_design_2022}
are critically needed 
to improve gait performance and injury models. \\[6.pt]
%%
%Computational models have also been used 
%to optimize the design of the shoe itself.
%%
%These models critically rely 
%on accurate constitutive modeling 
%of the elastomeric foam \cite{yang_design_2022}.
%%%%%%%%%%%%%%%%%%%%%%%%%%%%%%%%%%%%%%%%%%%%%%%%%%%%%%%%%%%%%%%%%%%%%%%%%%%%
{\sffamily{\bfseries{Mechanics of ultra-low density elastomeric foams.}}}
%%%%%%%%%%%%%%%%%%%%%%%%%%%%%%%%%%%%%%%%%%%%%%%%%%%%%%%%%%%%%%%%%%%%%%%%%%%%
To efficiently store mechanical energy during the stance phase 
and return it during toe-off, 
modern high-performance racing shoes employ
ultra-low-density elastomeric foams. 
Elastomers can undergo large, reversible deformations
that enable substantial elastic energy storage, 
while maintaining mechanical robustness under repeated loading \cite{treloar_rubber_1975}.
Their porous microstructure 
reduces mass and 
increases compliance, 
both of which are critical in endurance running.
Under uniaxial compression, 
elastomeric foams exhibit 
a characteristic nonlinear stress–stretch response: 
an initial linear elastic regime associated with reversible cell wall bending; 
a plateau regime associated with cell wall buckling; and 
a densification regime associated with pore collapse and rapidly increasing stiffness
%and the solid matrix takes over the load 
\cite{gibson_cellular_2003}. 
This S-shaped mechanical response 
enables a superior energy return 
at relatively low stresses 
when deformations remain below the densification regime.
At the same time, 
peak performance benefits of 
modern high-performance racing shoes
are typically limited to 150–300 miles of use \cite{rodrigocarranza_midsole_2024}.
While carbon-fiber plates 
exhibit minimal structural degradation, 
the ultra-low density polymeric foam 
undergoes viscoelastic fatigue under cyclic loading, 
leading to a progressive loss of resilience and energy return
\cite{aimar_compression_2024}.
It is becoming increasingly clear that
this limited lifespan 
is governed by midsole foam mechanics, 
rather than carbon-fiber plate integrity:
Designing the midsole foam is a trade off 
between performance, stability, and durability.\\[6.pt]
%%%%%%%%%%%%%%%%%%%%%%%%%%%%%%%%%%%%%%%%%%%%%%%%%%%%%%%%%%%%%%%%%%%%%%%%%%%%
{\sffamily{\bfseries{Dual-foam midsole architectures.}}}
%%%%%%%%%%%%%%%%%%%%%%%%%%%%%%%%%%%%%%%%%%%%%%%%%%%%%%%%%%%%%%%%%%%%%%%%%%%%
Not surprisingly, 
the mechanical properties of ultra-low-density polymeric foams 
are closely guarded and largely proprietary. 
Design laboratories 
fine-tune midsole foams by modulating 
polymer chemistry, 
void fraction, and 
foam microstructure, 
including pore size, pore shape, and the presence of microfillers \cite{aimar_compression_2024}.
As an illustrative example, 
Asics employs two proprietary midsole foams 
in its top-of-the-line racing shoe, 
\leap and \turbo. 
Notably, 
Asics is the only major manufacturer 
that offers two variants of its racing shoe
tailored to different running styles: 
the Metaspeed \edge, 
optimized for cadence, and 
the Metaspeed \sky, 
optimized for stride-length \cite{joubert_comparison_2021}. 
The primary distinction between these designs 
is the vertical arrangement of the two foams: 
In the \edge, 
\leap sits above the carbon-fiber plate and \turbo below, 
while in the \sky, this configuration is reversed.
However, 
because the mechanical properties of these foams 
are not publicly disclosed, 
it remains unclear to which extent 
this dual-foam arrangement 
truly provides a functional mechanical advantage. \\[6.pt]
{\sffamily{\bfseries{Automated model discovery.}}}
%%%%%%%%%%%%%%%%%%%%%%%%%%%%%%%%%%%%%%%%%%%%%%%%%%%%%%%%%%%%%%%%%%%%%%%%%%%%
Ultra-lightweight polymeric foams 
have undergone extensive 
experimental and theoretical study over several decades, 
resulting in a broad range of constitutive models 
that span cellular solids theory, hyperelasticity, and viscoelasticity \cite{gibson_1997}. 
Despite this progress, 
the selection of the appropriate foam models 
remains 
based on user experience and personal preference, 
rather than objective selection criteria.
Constitutive artificial neural networks 
enable the automated discovery of constitutive models, 
while simultaneously identifying optimal model parameters
\cite{linka_new_2023}. 
This deep learning framework 
identifies hyperelastic constitutive models for 
isotropic \cite{linka_brain_2023},  
transversely isotropic \cite{linka_automated_2023}, and
orthotropic \cite{martonova_automated_2024} materials, 
and extends naturally to 
viscoelastic \cite{holthusen_theory_2024} and 
inelastic material behavior \cite{holthusen_jax_2026}. 
These approaches apply across a wide range of materials, 
including 
biological tissues \cite{st_pierre_principal-stretch-based_2023}, 
foods \cite{st_pierre_mechanical_2024}, and 
synthetic materials \cite{mcculloch_automated_2024}.
While most existing studies focus on incompressible materials, 
recent work has introduced a bulk stiffness contributions 
into the constitutive neural network \cite{peirlinck_democratizing_2025}. 
Other machine-learning approaches 
also address the constitutive behavior
of elastomeric foams \cite{liang_neural_2008}. 
However, 
these vanilla type neural networks 
omit physics-based constraints in both 
the network architecture and 
the loss function and 
fail to leverage the extensive theoretical foundation 
of continuum mechanics and constitutive modeling.
Here we use constitutive neural networks
to discover physic-based models 
for ultra-low density foams
used in high-performance racing shoes.
%%%%%%%%%%%%%%%%%%%%%%%%%%%%%%%%%%%%%%%%%%%%%%%%%%%%%%%%%%%%%%%%%%%%%%%%%%%%
\section{{\sffamily{\bfseries{Methods}}}}
\label{sec:methods}
%%%%%%%%%%%%%%%%%%%%%%%%%%%%%%%%%%%%%%%%%%%%%%%%%%%%%%%%%%%%%%%%%%%%%%%%%%%%
In this study, we characterize 
two ultra-low density elastomeric polymeric foams
%%%%%%%%%%%%%%%%%%%%%%%%%%%%%%%%%%%%%%%%%%%%%%%%%%%%%%%%%%%%%%%%%%%%%%%%%%%%
\begin{figure}[b]
    \centering
    \includegraphics[width=1.0\linewidth,height=0.531\linewidth]{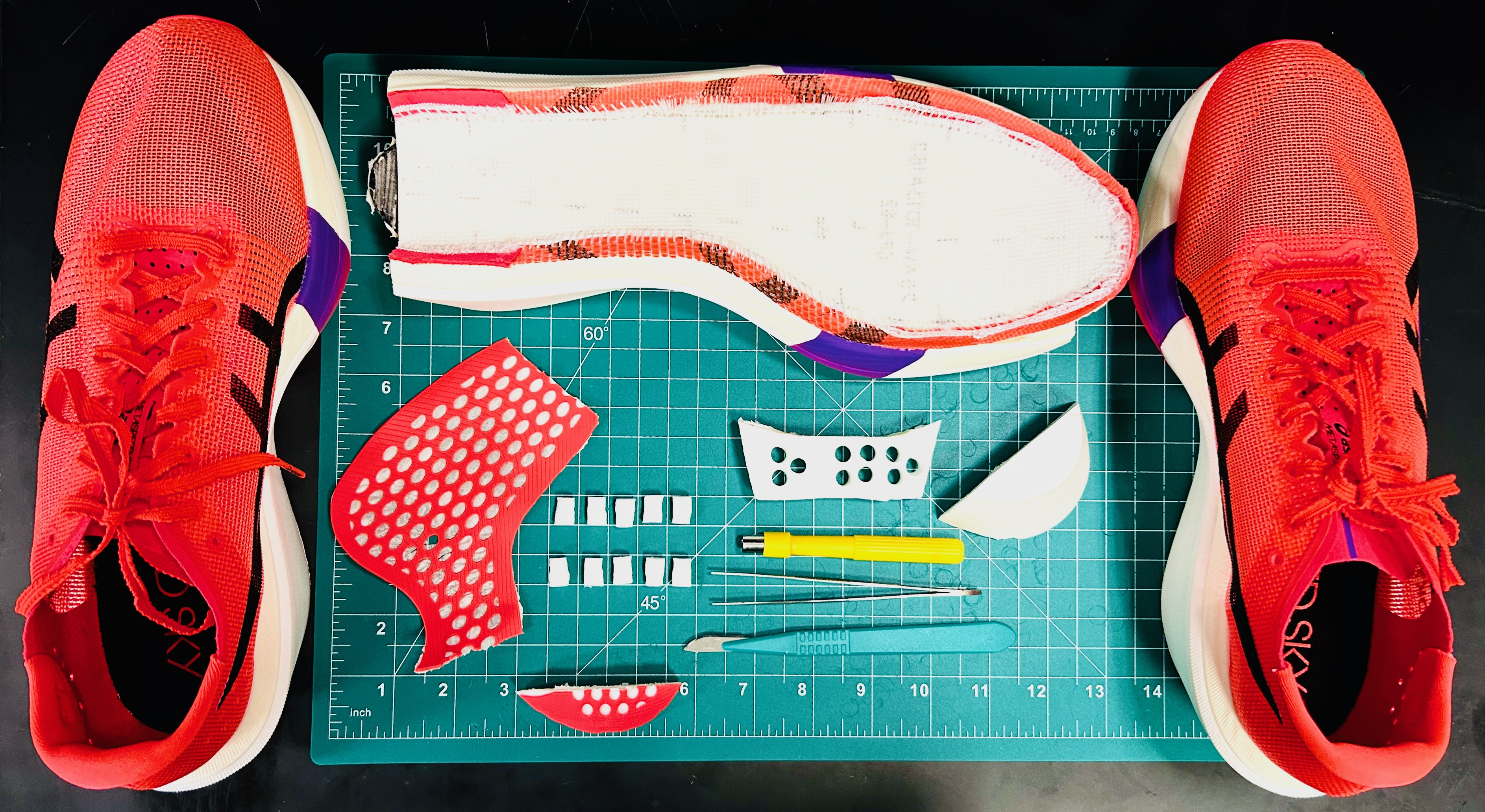}
    \caption{{\sffamily{\bfseries{Sample preparation.}}} 
We prepare samples from the 
ASICS Metaspeed \sky and \edge racing shoes 
by removing the rubber outsole 
and separating the carbon-fiber plate 
from the underlying foam. 
We section the extracted foam into slabs, 
and cut rectangular 
%5\,mm thick, 10\,mm wide, and 50\,mm long 
samples for uniaxial tension testing
and cylindrical 
%8\,mm diameter 10\,mm high 
samples for unconfined and confined compression, and shear testing. 
The images illustrate shoe disassembly, 
foam sectioning, and
specimen fabrication.}       
    \label{fig:samples}
\end{figure}
%%%%%%%%%%%%%%%%%%%%%%%%%%%%%%%%%%%%%%%%%%%%%%%%%%%%%%%%%%%%%%%%%%%%%%%%%%%%
%%%%%%%%%%%%%%%%%%%%%%%%%%%%%%%%%%%%%%%%%%%%%%%%%%%%%%%%%%%%%%%%%%%%%%%%%%%%
\begin{figure}[b]
    \centering
    \includegraphics[width=1.0\linewidth]{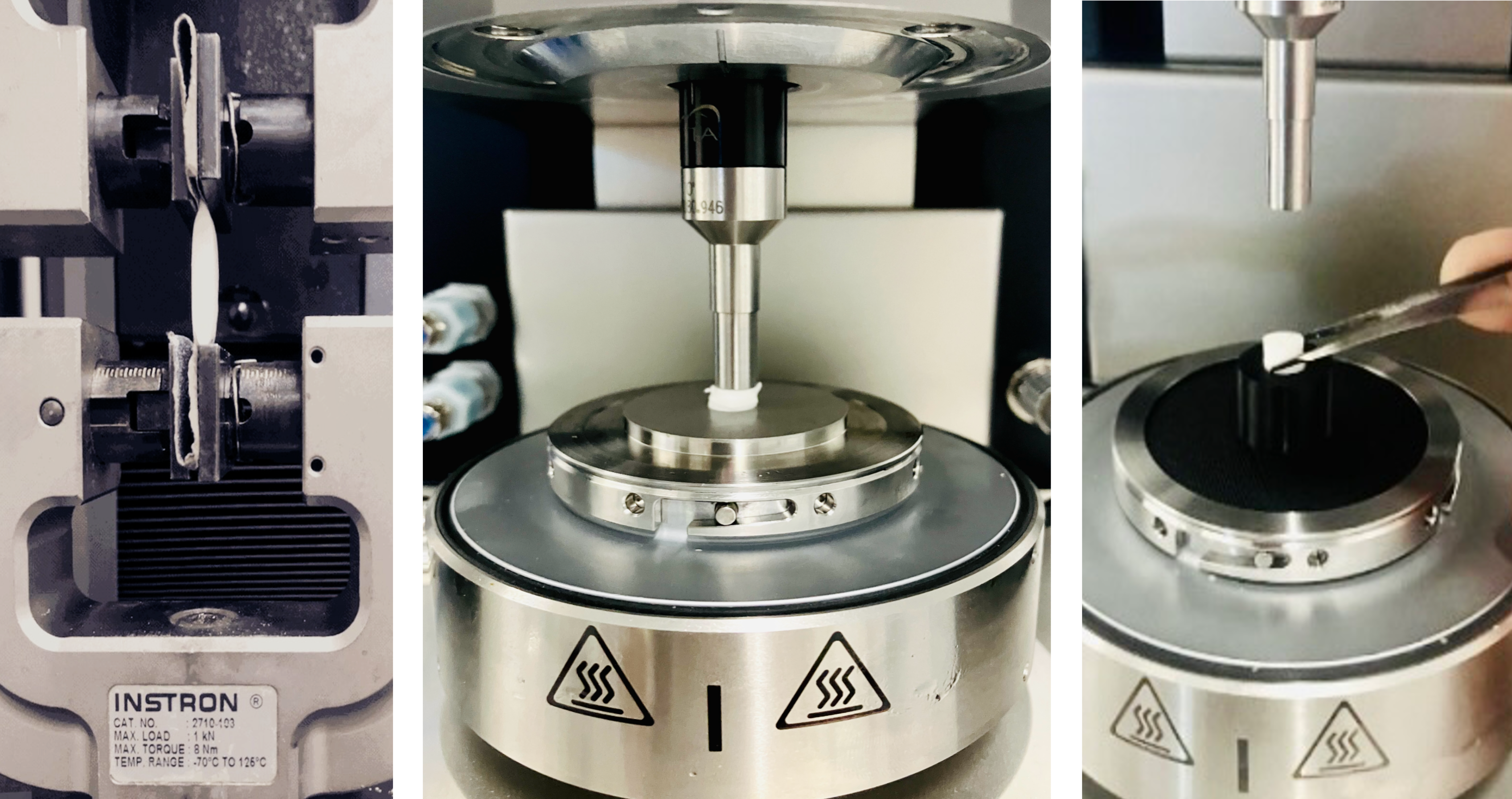}
    \caption{{\sffamily{\bfseries{Mechanical testing.}}} 
We test five samples of each foam in 
uniaxial tension, left, 
unconfined compression, middle, and 
confined compression, right.}       
    \label{fig:tests}
\end{figure}
%%%%%%%%%%%%%%%%%%%%%%%%%%%%%%%%%%%%%%%%%%%%%%%%%%%%%%%%%%%%%%%%%%%%%%%%%%%%
the \turbo and \leap 
used in the Asics Metaspeed
carbon-fiber-plated racing shoes. 
We test both foams in 
uniaxial tension, unconfined and confined compression, and simple shear, 
and use the experimental data 
to discover 
hyperelastic constitutive models for the foams
using a customized isotropic constitutive neural network. %\\[6.pt]
%%%%%%%%%%%%%%%%%%%%%%%%%%%%%%%%%%%%%%%%%%%%%%%%%%%%%%%%%%%%%%%%%%%%%%%%%%%%
\subsection{\sffamily{\bfseries{Sample preparation.}}}
%%%%%%%%%%%%%%%%%%%%%%%%%%%%%%%%%%%%%%%%%%%%%%%%%%%%%%%%%%%%%%%%%%%%%%%%%%%%
We extract samples from Asics 
Metaspeed \sky and \edge racing shoes. 
To start, 
we remove the rubber outer sole from each shoe 
by cutting with a surgical blade and 
gently pulling on the rubber. 
Then, 
we use a hot wire foam cutter (YaeTek Micromot Hot Wire ThermoCut Foam Cutting Machine, Yaemart, Duluth, GA) 
to cut along the bottom side of the carbon-fiber plate 
and separate it from the foam below. 
We then cut this piece of foam 
into 10 mm slabs along the length of the shoe. 
For the uniaxial tension tests, 
we cut rectangular samples of 
5\,mm thickness, 10\,mm width, and 50\,mm length.  
For the compression, shear, and confined compression tests,
we create cylindrical samples of 8\,mm diameter and 10\,mm height 
using a biopsy punch. 
Before testing, 
we measure the exact dimensions of each samples using calipers.
Figure \ref{fig:samples} shows details of the sample preparation. 
%\\[6.pt]
%%%%%%%%%%%%%%%%%%%%%%%%%%%%%%%%%%%%%%%%%%%%%%%%%%%%%%%%%%%%%%%%%%%%%%%%%%%%
%{\sffamily{\bfseries{Sample preparation.}}}
%%%%%%%%%%%%%%%%%%%%%%%%%%%%%%%%%%%%%%%%%%%%%%%%%%%%%%%%%%%%%%%%%%%%%%%%%%%%
\subsection{{\sffamily{\bfseries{Mechanical testing}}}}
\label{sec:mech-test-methods}
%%%%%%%%%%%%%%%%%%%%%%%%%%%%%%%%%%%%%%%%%%%%%%%%%%%%%%%%%%%%%%%%%%%%%%%%%%%%
We test 
$n=5$ samples of the \leap and 
$n=5$ samples of the \turbo
in uniaxial tension, unconfined and confined compression, and simple shear.
Figure \ref{fig:tests} shows the experimental test setups
for uniaxial tension, 
unconfined compression and simple shear, and 
confined compression,
from left to right.
\\[6.pt]
%%%%%%%%%%%%%%%%%%%%%%%%%%%%%%%%%%%%%%%%%%%%%%%%%%%%%%%%%%%%%%%%%%%%%%%%%%%%
{{\sffamily{\bfseries{Uniaxial tension.}}}}
%\label{sec:testUT}
%%%%%%%%%%%%%%%%%%%%%%%%%%%%%%%%%%%%%%%%%%%%%%%%%%%%%%%%%%%%%%%%%%%%%%%%%%%%
For the uniaxial tension tests,
we use an Instron 5848 
(Instron, Canton, MA) 
equipped with a 100\,N load cell. 
We tightly clamp 15\,mm length 
at each end of the sample 
using grips with sandpaper, 
to leave a gauge length of approximately 20\,mm. 
We zero the load cell to the weight of the foam sample, 
and then apply a pre-strain of 0.1,
corresponding to a pre-load of approximately 2\,kN, 
by gradually increasing the deformation 
until the axial force stabilizes above 0.1\,N. 
After preloading, 
we cyclically load the sample 
to a maximum stretch of $\lambda = 1.6$ 
at a rate of $\dot{\lambda} = 0.25$/s for six cycles; 
we then cyclically load the sample 
to a maximum stretch of $\lambda = 3.0$ 
at the same strain rate for six cycles; 
finally, 
we load the sample to failure 
at the same strain rate. \\[6.pt]
%%%%%%%%%%%%%%%%%%%%%%%%%%%%%%%%%%%%%%%%%%%%%%%%%%%%%%%%%%%%%%%%%%%%%%%%%%%%
{{\sffamily{\bfseries{Unconfined and confined compression.}}}}
%\label{sec:uniax-comp}
%%%%%%%%%%%%%%%%%%%%%%%%%%%%%%%%%%%%%%%%%%%%%%%%%%%%%%%%%%%%%%%%%%%%%%%%%%%%
For the unconfined compression tests, 
we use a HR20 discovery hybrid rheometer 
(TA Instruments, New Castle, DE) 
equipped with a 10\,N load cell and an 8mm tool. 
We again apply a pre-strain of 0.1,
corresponding to a pre-load of approximately 2\,kN,
by lowering the end effector 50\,$\mu$m at a time 
until the axial force stabilizes above 0.1\,N. 
After preloading, 
we cyclically load the sample 
to a minimum stretch of $\lambda = 0.4$ 
at a rate of $\dot{\lambda} = 0.25$/s for six cycles; 
and then compress the sample 
to the maximum force of the load cell of approximately 9\,N 
at the same strain rate. 
For the confined compression tests, 
we mount the cylindrical sample 
into a custom base plate 
with an 8\,mm inner diameter 
confined compression chamber,
and following the same testing protocol. \\[6.pt]
%%%%%%%%%%%%%%%%%%%%%%%%%%%%%%%%%%%%%%%%%%%%%%%%%%%%%%%%%%%%%%%%%%%%%%%%%%%%
{\sffamily{\bfseries{Simple shear.}}}
%%%%%%%%%%%%%%%%%%%%%%%%%%%%%%%%%%%%%%%%%%%%%%%%%%%%%%%%%%%%%%%%%%%%%%%%%%%%
For the shear tests, 
we use a HR20 discovery hybrid rheometer 
(TA Instruments, New Castle, DE) 
equipped with 
a custom 3D printed end effector and a platform 
with sandpaper glued to both. 
The sandpaper minimizes slippage 
between the sample and the device 
without using of an adhesive, 
which may alter the foam properties. 
First, we apply a preload 
similar to the case of unconfined compression.  
Then, we compress the samples 
to a stretch of $\lambda = 0.8$ 
at a rate of $\dot{\lambda} = 0.25$/s. 
Then, we apply a sinusoidal angular displacement 
with a maximum shear strain of $\gamma = 0.25$ 
and a maximum shear strain rate of $\dot{\gamma} = 0.25$/s, 
which corresponds to an angular frequency of 1\,rad/s. 
%%%%%%%%%%%%%%%%%%%%%%%%%%%%%%%%%%%%%%%%%%%%%%%%%%%%%%%%%%%%%%%%%%%%%%%%%%%%
\subsection{{\sffamily{\bfseries{Data processing}}}}
\label{sec:data_proc}% Move to appendix? not very interesting
%%%%%%%%%%%%%%%%%%%%%%%%%%%%%%%%%%%%%%%%%%%%%%%%%%%%%%%%%%%%%%%%%%%%%%%%%%%%
For each of the four mechanical testing modes 
in Section \ref{sec:mech-test-methods}, 
for each foam type,
we aim to derive a single stress-stretch curve,
which represents the average hyperelastic response of that foam. 
We convert the raw load-displacement recordings 
into stretch-stress measurements, 
average the loading and unloading curves for each sample, and 
then average across all $n=5$ samples. 
The the following subsections 
outline the precise process for each loading mode. \\[6.pt]
%%%%%%%%%%%%%%%%%%%%%%%%%%%%%%%%%%%%%%%%%%%%%%%%%%%%%%%%%%%%%%%%%%%%%%%%%%%%
{\sffamily{\bfseries{Uniaxial tension.}}}
%%%%%%%%%%%%%%%%%%%%%%%%%%%%%%%%%%%%%%%%%%%%%%%%%%%%%%%%%%%%%%%%%%%%%%%%%%%%
For each sample, 
we measure the initial cross sectional area $A$ 
and the initial length $L$ between the clamps. 
Throughout the experiment, 
we record the axial force $F(t)$ 
and the displacement $u(t)$ over time $t$.
We convert both measurements 
into the axial stretch $\lambda$
and the Piola stress $P_{\rm{11}}$, 
\begin{equation}
\lambda = 1 + \frac{u}{L} 
\quad \mbox{and} \quad
P_{\rm{11}} = \frac{F}{A}\,.
\end{equation}
By visually observing the tension experiment
and inspecting the stress-stretch curve, 
we conclude that a part of the behavior is inelastic,
since the stress reaches zero during unloading 
at stretches significantly larger than one. 
For all further analyses,  
we only consider the stretch range of $1.0 \le \lambda \le 1.3$. 
To obtain the stretch-stress curves 
$P_{\rm{11}}(\lambda)$ for each sample,
we average 
the first loading 
and final unloading curves 
and zero the stresses 
such that $P_{\rm{11}}|_{\lambda=1} = 0$. 
From the $n=5$ individual curves, 
we compute the mean $P_{\rm{11}}(\lambda)$ 
and the standard deviation $\sigma_{P_{\rm{11}}}$ 
across all five samples. \\[6.pt]
%%%%%%%%%%%%%%%%%%%%%%%%%%%%%%%%%%%%%%%%%%%%%%%%%%%%%%%%%%%%%%%%%%%%%%%%%%%%
{\sffamily{\bfseries{Unconfined\,and\,confined\,compression.}}}
%%%%%%%%%%%%%%%%%%%%%%%%%%%%%%%%%%%%%%%%%%%%%%%%%%%%%%%%%%%%%%%%%%%%%%%%%%%%
For each sample, 
we measure the initial cross sectional area $A$
and the initial sample height $H$.
Throughout the experiment, 
we record the axial force $F(t)$ 
and the deformed height of the sample $h(t)$
over time $t$. 
We convert both measurements 
into the axial stretch $\lambda$
and the Piola stress $P_{\rm{11}}$, 
\begin{equation}
\lambda = \frac{h}{H}
\quad \mbox{and} \quad
P_{\rm{11}} = \frac{F}{A} \,.
\end{equation} 
%
%Next, 
%for each of the final five loading and five unloading curves, 
%we interpolate to find the function 
%$P_{xx}^j(\lambda_x)$ for $\lambda_x \in [0.4, 1.0]$, 
%where $j \in \{1, 2, ..., 10\}$ 
%indicates which loading or unloading curve is being sampled. 
For all further analyses,  
we consider the stretch range of $0.4 \le \lambda \le 1.0$. 
For each sample, 
we exclude the first loading and unloading curves, 
% to account for the mullens effect [CITE]. Finally, we 
average the remaining ten curves,
and zero the stresses 
such that $P_{\rm{11}}|_{\lambda=1} = 0$. 
From the $n=5$ individual curves, we compute 
the mean $P_{\rm{11}}(\lambda)$ and 
the standard deviation $\sigma_{P_{\rm{11}}}$ 
across all five samples. 
We use the same procedure for both 
unconfined and confined compression. \\[6.pt]
%%%%%%%%%%%%%%%%%%%%%%%%%%%%%%%%%%%%%%%%%%%%%%%%%%%%%%%%%%%%%%%%%%%%%%%%%%%%
{\sffamily{\bfseries{Simple shear.}}}
%%%%%%%%%%%%%%%%%%%%%%%%%%%%%%%%%%%%%%%%%%%%%%%%%%%%%%%%%%%%%%%%%%%%%%%%%%%%
For each sample, 
we measure the initial radius $R$
and the initial sample height $H$.
Throughout the experiment, 
we record 
the torsion angle $\phi(t)$ and
the torque $T(t)$  
over time $t$. 
We convert both measurements into the shear strain $\gamma$
and the Piola stress $P_{\rm{12}}$,
\beq
\gamma = \frac{R}{H} \, \phi  %= F_{12}
\quad \mbox{and} \quad
{P}_{\rm{12}} = \frac{2}{\pi R^3} \, T \,,
\eeq
where we calculate the torque $T$ 
by integrating the shear stress $P_{\rm{12}}$
times its moment arm $r$ across the cross section,
${\rm{d}}A = r \, {\rm{d}}r \, {\rm{d}} \theta$, of the sample,
$ T 
%= \int_0^{2\pi} \int_0^{R} 
%  P_{\rm{xy}} \, r^2 {\rm{d}}r \, {\rm{d}}\theta
= 2\pi \int_{r=0}^{R} P_{\rm{12}} \,r^2 \, {\rm{d}}r$,
assume the following explicit shear stress-strain relation, 
$P_{\rm{12}} 
= 2\, 
[\, \partial \psi / \partial I_1
 +  \partial \psi / \partial I_2  \,] \, \gamma$,
and apply the trapezoidal rule,
$\int_{r=0}^R f(r) {\rm{d}}r 
\approx R \, [f(0) + f(R)]/2$,
to numerically approximate the integral 
\cite{stpierre_mechanics_meat_2023}.
These equations are exact for linear materials, 
%including the popular neo Hooke, Blatz Ko, or Mooney Rivlin models,
but they may very well under- or over-estimate 
the shear stress ${P}_{\rm{12}}$
for nonlinear hyperelastic materials 
depending on whether the shear stress-stretch curve is concave or convex.  
For all further analyses, 
we consider shear stretches, $\gamma \leq 0.2$,
for which the shear response is close to linear.
From the $n=5$ individual curves, 
we compute the mean
$P_{\rm{12}}(\gamma)$ and the standard deviation
$\sigma_{P_{\rm{12}}}$ across all five samples. \\[6.pt]
%%%%%%%%%%%%%%%%%%%%%%%%%%%%%%%%%%%%%%%%%%%%%%%%%%%%%%%%%%%%%%%%%%%%%%%%%%%%
{{\sffamily{\bfseries{Linear elastic stiffness and energy return.}}}}
%%%%%%%%%%%%%%%%%%%%%%%%%%%%%%%%%%%%%%%%%%%%%%%%%%%%%%%%%%%%%%%%%%%%%%%%%%%%
Using the stress-stretch pairs 
of the tension, compression, and shear experiments
in Figures \ref{fig:raw_data_leap} and \ref{fig:raw_data_turbo},
for relative deformations up to $10\%$,
we perform a linear regression 
to estimate the linear elastic stiffness in 
tension
${\textsf{E}}_{\rm{ten}} 
= \ten{\epsilon}_{\rm{ten}} \cdot \ten{\sigma}_{\rm{ten}} 
/ \ten{\epsilon}_{\rm{ten}} \cdot \ten{\epsilon}_{\rm{ten}}$,
compression
${\textsf{E}}_{\rm{com}} 
= \ten{\epsilon}_{\rm{com}} \cdot \ten{\sigma}_{\rm{com}} 
/ \ten{\epsilon}_{\rm{com}} \cdot \ten{\epsilon}_{\rm{com}}$, and
shear
${\textsf{G}}_{\rm{shr}} 
= \ten{\gamma}_{\rm{shr}} \cdot \ten{\tau}_{\rm{shr}} 
/ \ten{\gamma}_{\rm{shr}} \cdot \ten{\gamma}_{\rm{shr}}$,
where 
$\ten{\epsilon}_{\rm{ten}}, \ten{\epsilon}_{\rm{com}}, \ten{\gamma}_{\rm{shr}}$
are vectors that contain the discrete strain values up to $10\%$ 
across all five foams and 
$\ten{\sigma}_{\rm{ten}}, \ten{\sigma}_{\rm{com}}, \ten{\tau}_{\rm{shr}}$
are vectors of the associated stresses.
Using the full loading and unloading curves 
of the tension, compression, and shear experiments
in Figures \ref{fig:raw_data_leap} and \ref{fig:raw_data_turbo},
across the entire range of deformation,
we calculate the relative energy return,
$\eta = {\textsf{E}}_{\rm{unload}} / {\textsf{E}}_{\rm{load}}$,
as the ratio between the areas 
under the unloading curve ${\textsf{E}}_{\rm{unload}}$ 
and loading curves ${\textsf{E}}_{\rm{load}}$.
We report the stiffnesses,
${\textsf{E}}_{\rm{ten}}, {\textsf{E}}_{\rm{com}},{\textsf{E}}_{\rm{shr}}$,
and the relativ energy returns,
$\eta_{\rm{ten}},\eta_{\rm{com}},\eta_{\rm{shr}}$,
for both foams as means $\pm$ standard deviations.
%%%%%%%%%%%%%%%%%%%%%%%%%%%%%%%%%%%%%%%%%%%%%%%%%%%%%%%%%%%%%%%%%%%%%%%%%%%%
\subsection{{\sffamily{\bfseries{Continuum Mechanics}}}}
\label{sec:cont-mech}
%%%%%%%%%%%%%%%%%%%%%%%%%%%%%%%%%%%%%%%%%%%%%%%%%%%%%%%%%%%%%%%%%%%%%%%%%%%%
%% TODO: correct this section after additional experiments
%%%%%%%%%%%%%%%%%%%%%%%%%%%%%%%%%%%%%%%%%%%%%%%%%%%%%%%%%%%%%%%%%%%%%%%%%%%%
{{\sffamily{\bfseries{Kinematics.}}}}
%%%%%%%%%%%%%%%%%%%%%%%%%%%%%%%%%%%%%%%%%%%%%%%%%%%%%%%%%%%%%%%%%%%%%%%%%%%%
When comparing the stress-stretch curves 
from the unconfined and confined compression tests, 
we do not observe significant differences.
We conclude that the effective Poisson's ratio 
of both foams is approximately zero. 
Therefore, 
we assume that, in uniaxial tension and compression, 
the transverse strain is zero. 
For this setup, deformation gradient becomes
\begin{equation}
\ten{F} = 
\left[ \; 
\begin{array}{c@{\hspace*{0.2cm}}c@{\hspace*{0.2cm}}c}
\lambda & \gamma & 0 \\
0 & \alpha & 0\\ 
0 & 0 & 1 
\end{array}
\; \right] ,
\end{equation}
where 
$\lambda$ is the axial stretch, 
$\alpha$ is the transverse stretch, and 
$\gamma$ is the shear strain. 
%For uniaxial tension and compression, 
%$\lambda = \lambda(t)$, 
%$\gamma = 0$, and 
%$\alpha = 1$.
%For shear, 
%$\gamma = \gamma(t)$
%$\lambda = 0.8$, and 
%$\alpha = 1$. 
Next, we compute the right Cauchy Green deformation tensor, 
\begin{equation}
\ten{C} = 
\left[ \; 
\begin{array}{c@{\hspace*{0.2cm}}c@{\hspace*{0.2cm}}c}
\lambda^2 & \lambda\gamma & 0 \\
\lambda\gamma & \alpha^2 + \gamma^2 & 0\\ 
0 & 0 & 1
\end{array}
\; \right] .
\end{equation}
%\vspace*{0.2cm}
%%%%%%%%%%%%%%%%%%%%%%%%%%%%%%%%%%%%%%%%%%%%%%%%%%%%%%%%%%%%%%%%%%%%%%%%%%%%
{\sffamily{\bfseries{Invariant-based strain energy.}}}
%\label{sec:ps-strain-energy}
%%%%%%%%%%%%%%%%%%%%%%%%%%%%%%%%%%%%%%%%%%%%%%%%%%%%%%%%%%%%%%%%%%%%%%%%%%%%
We introduce the set of invariants,
\begin{equation}
\begin{array}{lcl}
I_1 &=& \tr (\ten{C}) = 1 + \alpha^2 + \lambda^2 + \gamma^2 \\ [4.pt]
I_2 &=& \frac{1}{2}[\, I_1^2 - \ten{C}:\ten{C} \,] = \alpha^2 + \gamma^2 + \lambda^2 + \alpha^2\lambda^2 \\ [4.pt]
J   &=& \det (\ten{F}) = \lambda \, \alpha \,.
\end{array}    
\label{invariants}
\end{equation}
and assume a strain energy function 
for which the contributions 
of these three invariants are fully decoupled,
\beq
  \psi(I_1, I_2, J) 
= \psi_{I_1}(I_1) 
+ \psi_{I_2}(I_2) 
+ \psi_{J}(J) \,.
\label{energy_f}
\eeq
With the partial derivatives 
of the invariants (\ref{invariants}) with respect to 
$\lambda$, $\gamma$, $\alpha$, at $\alpha = 1$,
\begin{equation}
\begin{array}{lcl@{\hspace*{0.4cm}}lcl@{\hspace*{0.4cm}}lcl}
    \displaystyle{\frac{\partial I_1}{\partial \lambda}} 
&=& 2 \lambda
&   \displaystyle{\frac{\partial I_1}{\partial \gamma}} 
&=& 2 \gamma
&   \displaystyle{\frac{\partial I_1}{\partial \alpha}} 
&=& 2 \\ [6.pt]
    \displaystyle{\frac{\partial I_2}{\partial \lambda}}
&=& 4 \lambda
&   \displaystyle{\frac{\partial I_1}{\partial \gamma}} 
&=& 2 \gamma
&   \displaystyle{\frac{\partial I_1}{\partial \alpha}}
&=& 2 + 2\lambda^2 \\ [6.pt]
    \displaystyle{\frac{\partial J}{\partial \lambda}}
&=& 1
&   \displaystyle{\frac{\partial I_1}{\partial \gamma}}
&=& 0 
&   \displaystyle{\frac{\partial I_1}{\partial \alpha}}
&=& \lambda \,,
\end{array}
\end{equation}
we compute the components of the Piola stress,
\[
%\begin{equation}
\begin{array}{@{\hspace*{0.0cm}}
             l@{\hspace*{0.1cm}}c@{\hspace*{0.05cm}}
             l@{\hspace*{0.1cm}}c@{\hspace*{0.05cm}}
             l@{\hspace*{0.1cm}}c@{\hspace*{0.1cm}}l}
    P_{11} 
&=& \displaystyle{\frac{\partial \psi}{\partial F_{11}}} 
&=& \displaystyle{\frac{\partial \psi}{\partial \lambda}} 
&=& \displaystyle{2\lambda\frac{\partial \psi}{\partial I_1} + 4\lambda\frac{\partial \psi}{\partial I_2} + \frac{\partial \psi}{\partial J}} 
%\\ [8.pt]
\end{array}
\]
\begin{equation}
\begin{array}{@{\hspace*{0.0cm}}
             l@{\hspace*{0.1cm}}c@{\hspace*{0.05cm}}
             l@{\hspace*{0.1cm}}c@{\hspace*{0.05cm}}
             l@{\hspace*{0.1cm}}c@{\hspace*{0.1cm}}l}
%%%
   P_{12} 
&=& \displaystyle{\frac{\partial \psi}{\partial F_{12}}} 
&=& \displaystyle{\frac{\partial \psi}{\partial \gamma}} 
&=& \displaystyle{2\gamma\frac{\partial \psi}{\partial I_1} + 2\gamma\frac{\partial \psi}{\partial I_2}} \\ [8.pt]
%%%
   P_{22} 
&=& \displaystyle{\frac{\partial \psi}{\partial F_{22}}} 
&=& \displaystyle{\frac{\partial \psi}{\partial \alpha}} 
&=& \displaystyle{2\frac{\partial \psi}{\partial I_1} + [2 + 2\lambda^2]\frac{\partial \psi}{\partial I_2} + \lambda\frac{\partial \psi}{\partial J}} .
\end{array}
\label{eqn:p22}
\end{equation}
To ensure 
that the stresses vanish %, $\ten{P} = \ten{0}$,
in the reference configuration, for
$\lambda = 1$, $\gamma = 0$, $\alpha = 1$, 
we need to enforce that
$P_{11}\rvert_{\ten{F}=\ten{I}} = 0$, thus
\begin{equation}
2\frac{\partial \psi}{\partial I_1}\rvert_{\ten{F}=\ten{I}} + 4\frac{\partial \psi}{\partial I_2}\rvert_{\ten{F}=\ten{I}} + \frac{\partial \psi}{\partial J}\rvert_{\ten{F}=\ten{I}}
= 0 \,,
\label{eqn:zero_stress_cond}
\end{equation}
%$ 2{\partial \psi}/{\partial I_1}\rvert_{\ten{F}
%=\ten{I}} 
%+ 4{\partial \psi}/{\partial I_2}\rvert_{\ten{F}=\ten{I}} 
%+ {\partial \psi}/{\partial J}\rvert_{\ten{F}=\ten{I}}
%= 0$ 
while
$P_{12}\rvert_{\ten{F}=\ten{I}} = 0$ alway holds.
For the free energy function in equation (\ref{energy_f}),
the above requirement
reduces to the overly restrictive condition,
$ {\partial \psi}/{\partial I_1}|_{I_1=3} 
= {\partial \psi}/{\partial I_2}|_{I_1=3} 
= {\partial \psi}/{\partial J}|_{J=1} = 0$. \\[6.pt]
%\vspace*{0.2cm}
%%%%%%%%%%%%%%%%%%%%%%%%%%%%%%%%%%%%%%%%%%%%%%%%%%%%%%%%%%%%%%%%%%%%%%%%%%%%
{\sffamily{\bfseries{Isochoric-invariant-based strain energy.}}}
%\label{sec:ps-strain-energy}
%%%%%%%%%%%%%%%%%%%%%%%%%%%%%%%%%%%%%%%%%%%%%%%%%%%%%%%%%%%%%%%%%%%%%%%%%%%%
In\-stead of using the set of invariants (\ref{invariants}), 
we introduce the set of isochoric invariants,
\beq
 \bar{I_1} 
=\frac{I_1}{J^{2/3}}
 \; \mbox{and} \; 
 \bar{I_2} = \frac{I_2}{J^{4/3}}
 \; \mbox{and} \; 
 J = \rm{det} (\ten{F})\,,
\eeq 
and assume a free energy function,
\beq
 \psi(\bar{I}_1, \bar{I}_2, J) 
= \psi_{\bar{I}_1}(\bar{I_1}) 
+ \psi_{\bar{I}_2}(\bar{I_2}) 
+ \psi_{J} (J) \,.
\eeq 
Now, with 
%${\partial \psi}/{\partial I_1}\rvert_{\ten{F}=\ten{I}} 
${\partial \psi_{\bar{I}_1}}/{\partial \bar{I}_1}|_{\bar{I}_1=3}=0$ and
%${\partial \psi}/{\partial I_2}\rvert_{\ten{F}=\ten{I}} 
${\partial \psi_{\bar{I}_2}}/{\partial \bar{I}_2}|_{\bar{I}_2=3}=0$ and 
%${\partial \psi}/{\partial J}\rvert_{\ten{F}=\ten{I}} 
$-2{\partial \psi_{\bar{I}_1}}/{\partial \bar{I}_1}|_{\bar{I}_1=3}
 -4{\partial \psi_{\bar{I}_2}}/{\partial \bar{I}_2}|_{\bar{I}_2=3} 
 + {\partial \psi_{J}}/{\partial J}|_{J=1}=0$,
the zero stress condition in equation (\ref{eqn:zero_stress_cond}) reduces to 
\begin{equation}
\frac{\partial \psi_{J}}{\partial J}|_{J=1} = 0 \,,
\label{eqn:gJ_cond1}
\end{equation}
which puts no restrictions on the forms of 
$\psi_{\bar{I}_1}$ and $\psi_{\bar{I}_2}$. 
We also require that, 
in the limits of $J\to 0$ or $J\to \infty$, 
the energy becomes infinite, $\psi \to \infty $, thus
\begin{equation}
  \lim_{J\to 0}\partial \psi_{J}(J) 
= \lim_{J\to +\infty}\partial \psi_{J}(J) 
= +\infty \,.
\label{eqn:gJ_cond2}
\end{equation}
%%%%%%%%%%%%%%%%%%%%%%%%%%%%%%%%%%%%%%%%%%%%%%%%%%%%%%%%%%%%%%%%%%%%%%%%%%%%
{\sffamily{\bfseries{Principal-stretch-based strain energy.}}}
\label{sec:ps-strain-energy}
%%%%%%%%%%%%%%%%%%%%%%%%%%%%%%%%%%%%%%%%%%%%%%%%%%%%%%%%%%%%%%%%%%%%%%%%%%%%
It can be useful 
to express some or all of the terms in the strain energy function 
in terms of the principal stretches 
$\{\,\lambda_1,\lambda_2,\lambda_3\,\}$
\cite{ogden_large_1972}, 
which are singular values of the deformation gradient $\ten{F}$. 
Here we introduce a principal-stretch-based strain energy~\cite{valanis1967strain},
\beq
  \psi 
= f(\lambda_1)+f(\lambda_2)+f(\lambda_3)
= \sum_{i = 1}^3 f(\lambda_i) \,,
\eeq
which is isotropic and objective as 
$\ten{F} \cdot \ten{Q}$, 
$\ten{Q} \cdot \ten{F}$, and 
$\ten{F}$ 
all have the same singular values 
for any proper orthogonal tensor $\ten{Q}$.
We also require 
that for no deformation, 
$\ten{F} = \ten{I}$,
the strain energy is zero,    
$\psi|_{\ten{F}=\ten{I}} 
= 3\,f|_{\lambda_i=1} = 0$, and
the stress is zero, so 
$P_{\rm{11}}|_{\ten{F}=\ten{I}}
= f'|_{\lambda_i=1} = 0$, thus
$f|_{\lambda_i=1} = f'|_{\lambda_i=1} = 0$.\\[6.pt]
For the uniaxial tension or compression experiments,
with $\gamma = 0$, 
the principal stretches are 
$\lambda_1 = \lambda$ and 
$\lambda_2 = \alpha$ and 
$\lambda_3 = 1$,
the strain energy is 
$\psi = f(\lambda) + f(\alpha)$, and
the normal stresses are
\beq
  P_{\rm{11}} 
= \frac{\partial \psi}{\partial\lambda} = f'(\lambda) 
  \; \mbox{and} \;
  P_{\rm{22}} 
= \frac{\partial \psi}{\partial\alpha} = f'(\alpha) \,.
\eeq
In our case, 
with $\alpha = 1$, $P_{22} = f'(1) = 0$. \\[6.pt]
For the shear experiments, 
with $\alpha = 1$, 
the principal stretches are 
$\lambda_{1,2}^2 
= \frac{1}{2}\,[1 + \lambda^2 + \gamma^2] 
\pm \frac{1}{2}\,\sqrt{[ 1 + \lambda^2 + \gamma^2]^2 - 4\, \lambda^2}$,
their derivatives with respect to the shear strain $\gamma$ are
$ \partial\lambda_{1,2}/{\partial \gamma} 
= \frac{1}{2}\, \gamma
[\,1\pm[1 + \lambda^2 + \gamma^2]
    /\sqrt{[1 + \lambda^2 + \gamma^2]^2 - 4\lambda^2}
 \,] / \lambda_{1,2}$, and
the shear stress is
\begin{equation}
P_{12} =\frac{\partial\psi}{\partial \gamma} = f'(\lambda_1)\frac{\partial\lambda_1}{\partial \gamma} + f'(\lambda_2)\frac{\partial\lambda_2}{\partial \gamma} \,.
\end{equation}
%%%%%%%%%%%%%%%%%%%%%%%%%%%%%%%%%%%%%%%%%%%%%%%%%%%%%%%%%%%%%%%%%%%%%%%%%%%%
\subsection{{\sffamily{\bfseries{Automated model discovery}}}}
%%%%%%%%%%%%%%%%%%%%%%%%%%%%%%%%%%%%%%%%%%%%%%%%%%%%%%%%%%%%%%%%%%%%%%%%%%%%
To discover the best constitutive model and parameters  
for ultra-low density elastomeric foams, 
we create a constitutive neural network 
which takes 
the invariants $\{ \bar{I}_1, \bar{I}_2, J \}$, and 
the principal stretches $\{ \lambda_1, \lambda_2, \lambda_3 \}$ 
as inputs and outputs the strain energy function $\psi$
from which we derive the stress.
Specifically, we integrate 
single-invariant terms,
mixed-invariant terms, and
principal-stretch terms
in a single free energy function of the following form, 
\beq
  \psi 
= \psi_{\bar{I}_1}
+ \psi_{\bar{I}_2}
+ \psi_{J}
+ \psi_{\bar{I}_1,J}
+ \psi_{\bar{I}_2,J}
+ \psi_{\lambda_i} .
\label{psi_network}
\eeq
In the following, 
we briefly motivate and introduce these six terms.
\\[6.pt]
%%%%%%%%%%%%%%%%%%%%%%%%%%%%%%%%%%%%%%%%%%%%%%%%%%%%%%%%%%%%%%%%%%%%%%%%%%%%
{\sffamily{\bfseries{Single-invariant terms.}}}
%%%%%%%%%%%%%%%%%%%%%%%%%%%%%%%%%%%%%%%%%%%%%%%%%%%%%%%%%%%%%%%%%%%%%%%%%%%%
Previous work has shown 
how to build a constitutive neural network 
for isotropic incompressible materials 
with the invariants $\{ {I}_1, {I}_2 \}$ as input \cite{linka_new_2023}. 
Here, 
we use a similar architecture 
with linear, quadratic, linear exponential, and quadratic exponential 
activation functions, 
but now 
in terms of the invariant $\bar{I}_1$,
\beq
\begin{array}{@{\hspace*{0.05cm}}lcl@{\hspace*{0.05cm}}l@{\hspace*{0.0cm}}cl@{\hspace*{0.05cm}}l@{\hspace*{0.05cm}}l}
    \psi_{{\bar{I}_1}}
&=& w_1 & [\, \bar I_1 - 3 \,]
&+& w_2 & [\, \exp (w_2^*[\bar I_1 - 3]) &- 1 ] \\
&+& w_3 & [\, \bar I_1 - 3 \,]^2
&+& w_4 & [\, \exp (w_4^*[\bar I_1 - 3]^2) &- 1 ] ,
\end{array}
\label{eqn:psi_I1}
\eeq
and the invariant $\bar{I}_2$,
\beq
\begin{array}{@{\hspace*{0.05cm}}lcl@{\hspace*{0.05cm}}l@{\hspace*{0.0cm}}cl@{\hspace*{0.05cm}}l@{\hspace*{0.05cm}}l}
    \psi_{{\bar{I}_2}}
&+& w_5 & [\, \bar I_2 - 3 \,]
&+& w_6 & [\, \exp (w_6^*[\bar I_2 - 3]) &- 1 ] \\
&+& w_7 & [\, \bar I_2 - 3 \,]^2
&+& w_8 & [\, \exp (w_8^*[\bar I_2 - 3]^2) &- 1 ] .
\end{array}
\label{eqn:psi_I2}
\eeq
In addition, we introduce the new the function $\psi_{J}$
guided by previous work \cite{mcculloch_automated_2024},
to ensure polyconvexity when $J \neq 1$ \cite{boes_mechanics_2025},
while satisfying 
the zero stress condition (\ref{eqn:gJ_cond1}) and 
the limit conditions (\ref{eqn:gJ_cond2}),
\begin{equation}
\begin{array}{lcll@{\hspace*{0.1cm}}l}
    \psi_{J}
&=& w_9 & [\, J^{w_9^*} - w_9^*\ln(J)&-1\,]  \\
&+& w_{10} & [\, \exp(w_{10}^*(\ln (J))^2) &-1\,]\,.
\end{array}
\label{eqn:psi_J}
\end{equation}
Both terms have been previously featured in  
constitutive models for rubber-like materials
\cite{hencky_uber_1928, neff_exponentiated_2015, landauer_experimental_2019}. 
\\[6.pt]
%
%We integrate this expression with 
%the classical eight-term constitutive neural network
%\cite{linka_automated_2023},
%to obtain the following single-invariant free energy function,
%\beq
%\begin{array}{@{\hspace*{0.05cm}}lcl@{\hspace*{0.05cm}}l@{\hspace*{0.0cm}}cl@{\hspace*{0.05cm}}l@{\hspace*{0.05cm}}l}
%    \psi_{\rm{si}}%(I_1, I_2, J; w_i, w_i^*)
%&=& w_1 & [\, \bar I_1 - 3 \,]
%&+& w_2 & [\, \exp (w_2^*[\bar I_1 - 3]) &- 1 ] \\
%&+& w_3 & [\, \bar I_1 - 3 \,]^2
%&+& w_4 & [\, \exp (w_4^*[\bar I_1 - 3]^2) &- 1 ] \\
%&+& w_5 & [\, \bar I_2 - 3 \,]
%&+& w_6 & [\, \exp (w_6^*[\bar I_2 - 3]) &- 1 ] \\
%&+& w_7 & [\, \bar I_2 - 3 \,]^2
%&+& w_8 & [\, \exp (w_8^*[\bar I_2 - 3]^2) &- 1 ] \\
%&&&&+& w_9 & [\, J^{w_9^*} - w_9^*\ln (J) &- 1 ] \\
%&&&&+& w_{10} & [\, \exp (w_{10}^*[\ln (J))^2] &- 1 ]
%\end{array}
%\label{eqn:si-terms}
%\eeq
%%%%%%%%%%%%%%%%%%%%%%%%%%%%%%%%%%%%%%%%%%%%%%%%%%%%%%%%%%%%%%%%%%%%%%%%%%%%
{\sffamily{\bfseries{Mixed-invarint terms.}}}
%%%%%%%%%%%%%%%%%%%%%%%%%%%%%%%%%%%%%%%%%%%%%%%%%%%%%%%%%%%%%%%%%%%%%%%%%%%%
Each term in equations (\ref{eqn:psi_I1},\ref{eqn:psi_I2},\ref{eqn:psi_J})
only depends on a single invariant, $\{\bar{I_1}, \bar{I_2}, J\}$. 
However, for many foams, 
the volumetric and deviatoric responses 
are inherently coupled. 
For example,  
foams are often 
stiffer in tension than in compression 
due to their porous microstructure. 
%\cite{gupta_comparison_2010}. 
We incorporate this behavior
through the following mixed-invariant terms,
%\begin{align}
%\psi_{\bar{I}_1, \bar{I}_2, J} 
%&= w_{11}J^{w_{11}^*} [\bar{I_1} - 3] + w_{12}J^{w_{12}^*}[\bar{I_2} - 3] \,. \label{eqn:psi_I1_I2_J}
%\end{align}
\beq
\begin{array}{lcl}
\psi_{\bar{I}_1, J} 
&=& w_{11}\,J^{w_{11}^*} \; [\bar{I_1} - 3] \\
\psi_{\bar{I}_2, J} 
&=& w_{12}\,J^{w_{12}^*} \; [\bar{I_2} - 3] \,. 
\end{array}
\label{eqn:psi_I1_I2_J}
\eeq
Both terms take larger values in tension than in compression
for positive exponents, ${w_{11}^*},{w_{12}^*} > 0$,
and they satisfy the constraint condition (\ref{eqn:zero_stress_cond}).
However, both terms may violate the polyconvexity condition. 
In particular, for $w_{11}^* > \frac{2}{3}$ and for all values of $w_{12}^*$, 
the strain energies 
$\psi_{\bar{I}_1, J}$ and $\psi_{\bar{I}_2, J}$
are not polyconvex or even rank-one convex.   \\[6.pt]
%%%%%%%%%%%%%%%%%%%%%%%%%%%%%%%%%%%%%%%%%%%%%%%%%%%%%%%%%%%%%%%%%%%%%%%%%%%%
{\sffamily{\bfseries{Principal-stretch terms.}}}
%%%%%%%%%%%%%%%%%%%%%%%%%%%%%%%%%%%%%%%%%%%%%%%%%%%%%%%%%%%%%%%%%%%%%%%%%%%%
To address the challenge of polyconvexity of the mixed-invariant terms,
while still allowing for a notable tension-compression asymmetry,
we consult the classical Ogden-Hill foam model \cite{hill_aspects_1979}, 
with
$\psi_{\lambda_i,J}
=
\sum_{i=1}^{n}
 {2\,\mu_i}
[\, \lambda_1^{\alpha_i}
  + \lambda_2^{\alpha_i}
  + \lambda_3^{\alpha_i} - 3
  + [\, J^{-\alpha_i \, \beta_i} -1 ]/{\beta_i} ]/{\alpha_i^2}$.
For the special case with $\beta_i \to 0$,
we can express this model exclusively in terms of the
three principal stretches 
$\{ \lambda_1$, $\lambda_2$, $\lambda_3\}$ 
as outlined in Section \ref{sec:ps-strain-energy},
$ \psi_{\lambda_i}
= \sum_{i=1}^n {\mu_i}/{\alpha_i^2}
  \sum_{j=1}^3 [ \lambda_j^{\alpha_i} - \alpha_i \ln (\lambda_j) - 1]$.
Here we consider the first two terms of this series, $n = 2$, 
and obtain the following principal-stretch terms,
\beq
\begin{array}{lcl}
    \psi_{\lambda_i} 
&=& w_{13} \sum_{j=1}^3 
    [\,\lambda_j^{w_{13}^*} \!- w_{13}^* \, \ln (\lambda_j) -1\,] \\ [4.pt]
&+& w_{14} \sum_{j=1}^3 
    [\,\lambda_j^{w_{14}^*} \!- w_{14}^* \, \ln (\lambda_j) -1\,]\,,
\end{array}
\label{eqn:psi_lambda}
\eeq
which we include into the principal-stretch based part of our neural network
\cite{buganza_polyconvex_pann_2025,st_pierre_principal-stretch-based_2023}.
% A visual representation of the architecture of the neural network described in this section can be found in figure \ref{fig:cann-arch}, and a complete analytical expression for the strain energy in terms of the model weights and the invariants is shown in equation \ref{eqn:cann-no-cross-terms}.
\\[6.pt]
%%%%%%%%%%%%%%%%%%%%%%%%%%%%%%%%%%%%%%%%%%%%%%%%%%%%%%%%%%%%%%%%%%%%%%%%%%%%
{\sffamily{\bfseries{Constitutive neural network.}}}
%%%%%%%%%%%%%%%%%%%%%%%%%%%%%%%%%%%%%%%%%%%%%%%%%%%%%%%%%%%%%%%%%%%%%%%%%%%%
%%%%%%%%%%%%%%%%%%%%%%%%%%%%%%%%%%%%%%%%%%%%%%%%%%%%%%%%%%%%%%%%%%%%%%%%%%%%
\begin{figure}[t]
    \centering
    \includegraphics[width=1.0\linewidth]{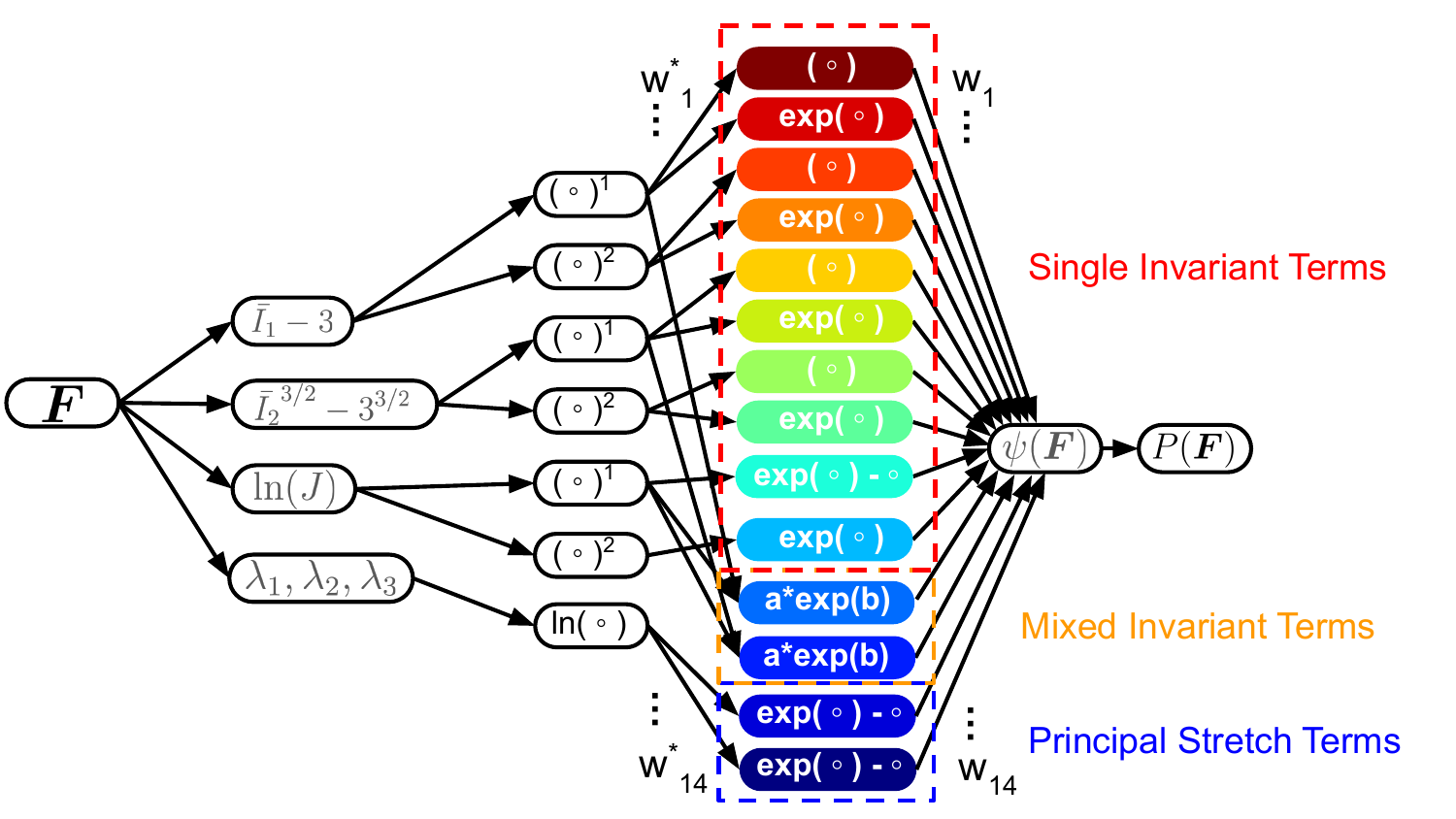}
    \caption{{\sffamily{\bfseries{Constitutive neural network for ultra-low density elastomeric foams.}}} The network takes sets of three invariants 
$\bar{I}_1, \bar{I}_2, J$ or three principal stretches
$\lambda_1,\lambda_2,\lambda_3$ as input
and learns a free energy function $\psi$ from which we derive the stress. 
The free energy function is made up of 14 terms, 
four first- and four second-invariant terms, $\psi_{\bar{I}_1}$ and $\psi_{\bar{I}_2}$,
two third-invariant terms, $\psi_{J}$,
two mixed-invariant terms, $\psi_{\bar{I}_1,J}$ and $\psi_{\bar{I}_2,J}$, and
two principal-stretch terms, $\psi_{\lambda_i}$.
During training, we selectively activate different combinations of terms
to discover the best model for two ultra-low density elastomeric foams.}       
%This figure shows all the terms that are used in the constitutive neural networks in this paper.     
%    The first set of terms are each a function of a single invariant, and have been used in previous constitutive neural networks \cite{linka_new_2023, boes_mechanics_2025} or are included in well established constitutive models  \cite{neff_exponentiated_2015, landauer_experimental_2019}. The second set of terms are each a function of multiple invariants. The final set of terms are a function of the principal stretches and both are a special case of the Ogden-Hill foam model\cite{hill_aspects_1979}. }
    \label{fig:cann-arch}
\end{figure}
%%%%%%%%%%%%%%%%%%%%%%%%%%%%%%%%%%%%%%%%%%%%%%%%%%%%%%%%%%%%%%%%%%%%%%%%%%%%
Figure \ref{fig:cann-arch} 
illustrates the complete network architecture 
for the free energy function (\ref{psi_network})
as a sum of the 14 individual terms from equations (\ref{eqn:psi_I1}) to (\ref{eqn:psi_lambda}),
four first- and four second-invariant terms, $\psi_{\bar{I}_1}$ and $\psi_{\bar{I}_2}$,
two third-invariant terms, $\psi_{J}$,
two mixed-invariant terms, $\psi_{\bar{I}_1,J}$ and $\psi_{\bar{I}_2,J}$, and
two principal-stretch terms, $\psi_{\lambda_i}$.
During automated model discovery, 
we include different subsets of these 14 terms 
into network training,
and compare the goodness of fit 
of the discovered models. 
Some of these models will include only polyconvex terms,
but we will also discover and discuss models, 
which feature non-polyconvex terms. \\[6.pt]
%%%%%%%%%%%%%%%%%%%%%%%%%%%%%%%%%%%%%%%%%%%%%%%%%%%%%%%%%%%%%%%%%%%%%%%%%%%%
{\sffamily{\bfseries{Polyconvexity.}}}
%%%%%%%%%%%%%%%%%%%%%%%%%%%%%%%%%%%%%%%%%%%%%%%%%%%%%%%%%%%%%%%%%%%%%%%%%%%%
A common requirement 
is to restrict the free energy function $\psi$ to be polyconvex, 
as this ensures 
the existence of at least one local minimum  
%for the associanted boundary value problems 
\cite{ball_convexity_1976}. 
Much research 
has focused on identifying polyconvex strain energy functions, 
and finding conditions 
under which particular strain energy functions are polyconvex \cite{hartmann_polyconvexity_2003}. 
However, 
microstructural materials such as foams 
might not always be well-represented by polyconvex functions \cite{suchocki_polyconvex_2021}, 
and recent work suggests 
that global polyconvexity 
may not be the best requirement 
to ensure a physically reasonable behavior 
\cite{wollner_search_2026}. 
In the network architecture in Figure \ref{fig:cann-arch}, 
the single invariant terms, 
$\psi_{\bar{I}_1}, \psi_{\bar{I}_2}, \psi_{J}$,
and 
the principal-stretch terms,
$\psi_{\lambda_i}$  
are polyconvex by design, 
provided that all weight $w_i^*$ and $w_i$ are non-negative. 
Yet, the mixed invariant terms
$\psi_{\bar{I}_1,J}, \psi_{\bar{I}_2,J}$
are not necessarily polyconvex,
even for non-negative weights $w_i$ and $w_i^*$,
and we may have to consult alternative solution strategies
tailored to non-convex multi-well potentials \cite{jones_lse_icnn_2025}.
\\[6.pt]
%%%%%%%%%%%%%%%%%%%%%%%%%%%%%%%%%%%%%%%%%%%%%%%%%%%%%%%%%%%%%%%%%%%%%%%%%%%%
\subsection{{\sffamily{\bfseries{Loss function}}}}
%%%%%%%%%%%%%%%%%%%%%%%%%%%%%%%%%%%%%%%%%%%%%%%%%%%%%%%%%%%%%%%%%%%%%%%%%%%%
%Using the continuum mechanics equations in section \ref{sec:cont-mech} and the constitutive neural network architecture proposed in sections \ref{sec:cann-arch}, it is possible to compute the axial Piola stress $\hat{P}_{11}(\lambda, \gamma)$, the shear Piola stress $\hat{P}_{12}(\lambda, \gamma)$, and the transverse Piola stress $\hat{P}_{22}(\lambda, \gamma)$ predicted by the model for a given value of stretch $\lambda$ and shear strain $\gamma$. For each torsion experiment, we have a set of measured strains $\gamma^{(i)}$ and stresses $P_{12}^{(i)}$. For each tension or compression experiment, we have a set of measured stretches $\lambda^{(i)}$ and axial stresses $P_{11}^{(i)}$; furthermore, we assume that the transverse stress $P_{22}^{(i)} = 0$ due to the plane stress boundary conditions. In order 
To train the constitutive neural network in Figure~\ref{fig:cann-arch},
we minimize a loss function
that consists of 
a weighted squared error between the 
model stresses $P(\lambda_i)$ and 
the experimental stresses $P_i$,
and add an $\textsf{L}_{0.5}$ regularization term 
that promotes sparsity of the model \cite{frank_1993}, 
\beq
\begin{array}{ @{\hspace*{0.00cm}}
              l@{\hspace*{0.05cm}}c@{\hspace*{0.05cm}}
              l@{\hspace*{-.02cm}}l@{\hspace*{0.03cm}}
              l@{\hspace*{0.00cm}}l@{\hspace*{0.05cm}}
              l@{\hspace*{0.02cm}}l}    
    \textsf{L} 
&=& \sum_{i=1}^{n_{\rm{ten}}} 
&[& [\, P_{11}(\lambda_i) - P_{11,i} \,]^2 
&+& P_{22}(\lambda_i)^2]
  & / \, (P_{11,i}^{\rm{max}})^2\\
%&+& \sum_{i=1}^{n_{\rm{ten}}}  
%    [\, P_{22}(\lambda_i) - 0 \,]^2  
%  & / \; (P_{11,i}^{\rm{max}})^2\\
&=& \sum_{i=1}^{n_{\rm{com}}} 
&[& [\, P_{11}(\lambda_i) - P_{11,i} \,]^2 
&+& P_{22}(\lambda_i)^2]   
  & / \, (P_{11,i}^{\rm{min\,}})^2\\
%&+& \sum_{i=1}^{n_{\rm{com}}}  
%    [\, P_{22}(\lambda_i) - 0 \,]^2  
%  & / (P_{11,i}^{\rm{max}})^2\\
&+& \sum_{i=1}^{n_{\rm{shr}}} 
&[& [\, P_{12}(\gamma_i) - P_{12,i} \,]^2  
&&&/ \,(P_{12,i}^{\rm{max}})^2 \\
%+ \alpha 
%    \sum_{i=1}^{n_{w}}|\, w_i \,|^{0.5}
%  \\
&+& \sum_{i=1}^{n_{w}}
& & |\,w_i \,|^{0.5} \; \alpha
\end{array}
\label{eqn:loss}
\eeq
%There are multiple possible ways to define the weights, but we choose to give equivalent weight to tension, compression, and shear by choosing 
%\begin{equation}
%    W_{shr}^{(i)} = \frac{1}{\max\limits_j \left(P_{12}^{(j)}\right)^2}
%\end{equation}
%\begin{equation}
%W_{ax}^{(i)} = W_{trans}^{(i)} = \begin{cases} 
%      \frac{1}{\max\limits_{\lambda_j>1} \left(P_{11}^{(j)}\right)^2}, & \lambda^{(i)} \geq 1 \\
%      \frac{1}{\max\limits_{\lambda_j<1} \left(P_{11}^{(j)}\right)^2}, & \lambda^{(i)} < 1 \\
%   \end{cases}
%\end{equation}
First, we train without the regularization term
by setting the regularization parameter to zero,
$\alpha = 0$.
Then, 
we initialize the network 
with the learned weights, 
and systematically vary the regularization parameter $\alpha$ \cite{mcculloch_sparse_2024}.
We select the model that has the fewest terms 
without significantly sacrificing performance 
relative to the model trained with $\alpha = 0$. 
%%%%%%%%%%%%%%%%%%%%%%%%%%%%%%%%%%%%%%%%%%%%%%%%%%%%%%%%%%%%%%%%%%%%%%%%%%%%
\section{{\sffamily{\bfseries{Results}}}}
\label{sec:results}
%%%%%%%%%%%%%%%%%%%%%%%%%%%%%%%%%%%%%%%%%%%%%%%%%%%%%%%%%%%%%%%%%%%%%%%%%%%%
\subsection{{\sffamily{\bfseries{Mechanical testing}}}}
\begin{figure*}[h!]
    \centering
    \includegraphics[width=0.85\linewidth]{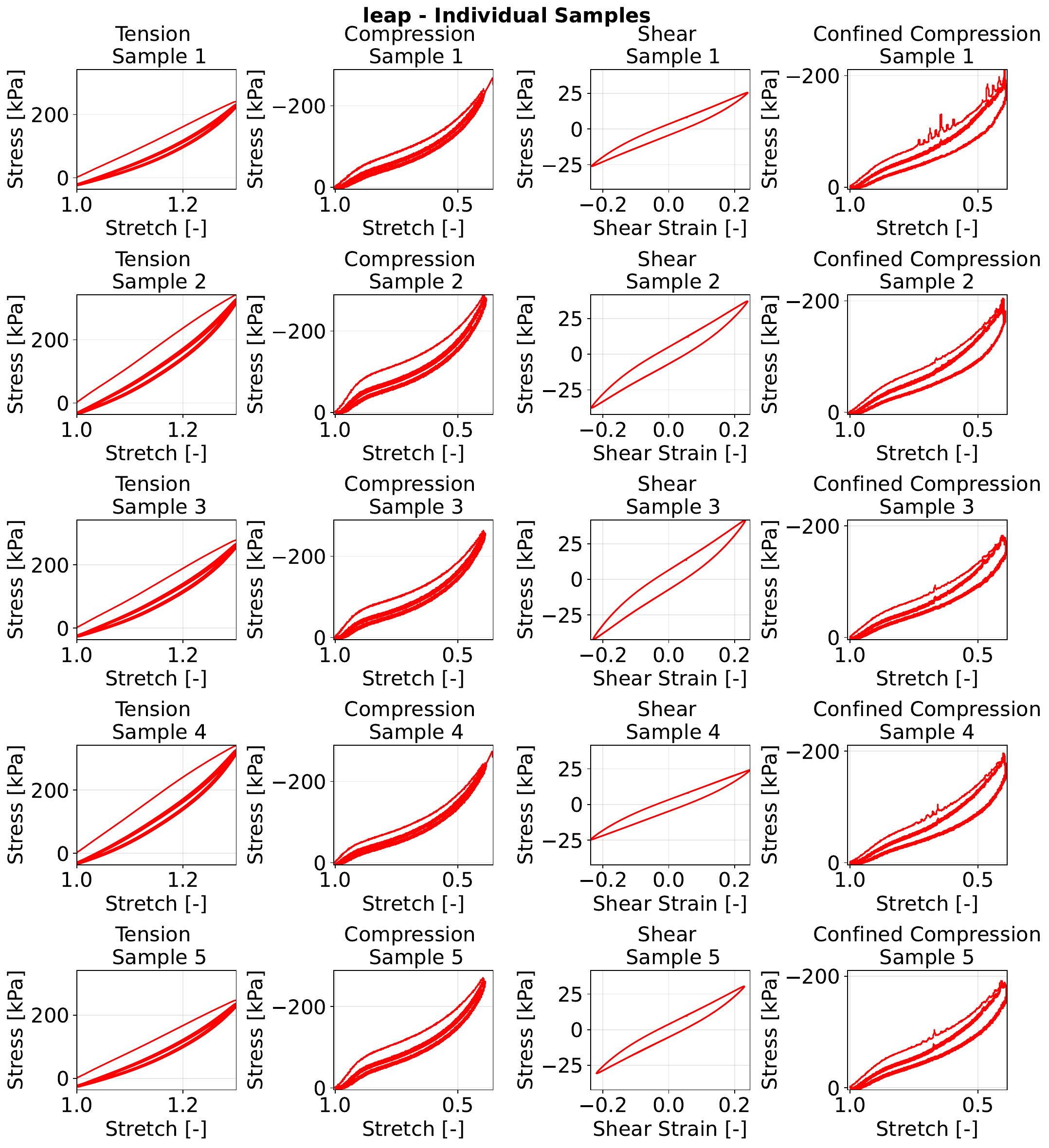}
    \caption{{\sffamily{\bfseries{\leap raw data from uniaxial tension, unconfined compression, simple shear, and confined compression experiments.}}}
Piola stress vs stretch or shear strain 
measurements for all \leap samples. 
Columns correspond to the experimental setups, 
tension, compression, shear, and confined compression;
rows corresponds to the different samples.}             
\label{fig:raw_data_leap}
\end{figure*}

\begin{figure*}[h!]
    \centering
    \includegraphics[width=0.85\linewidth]{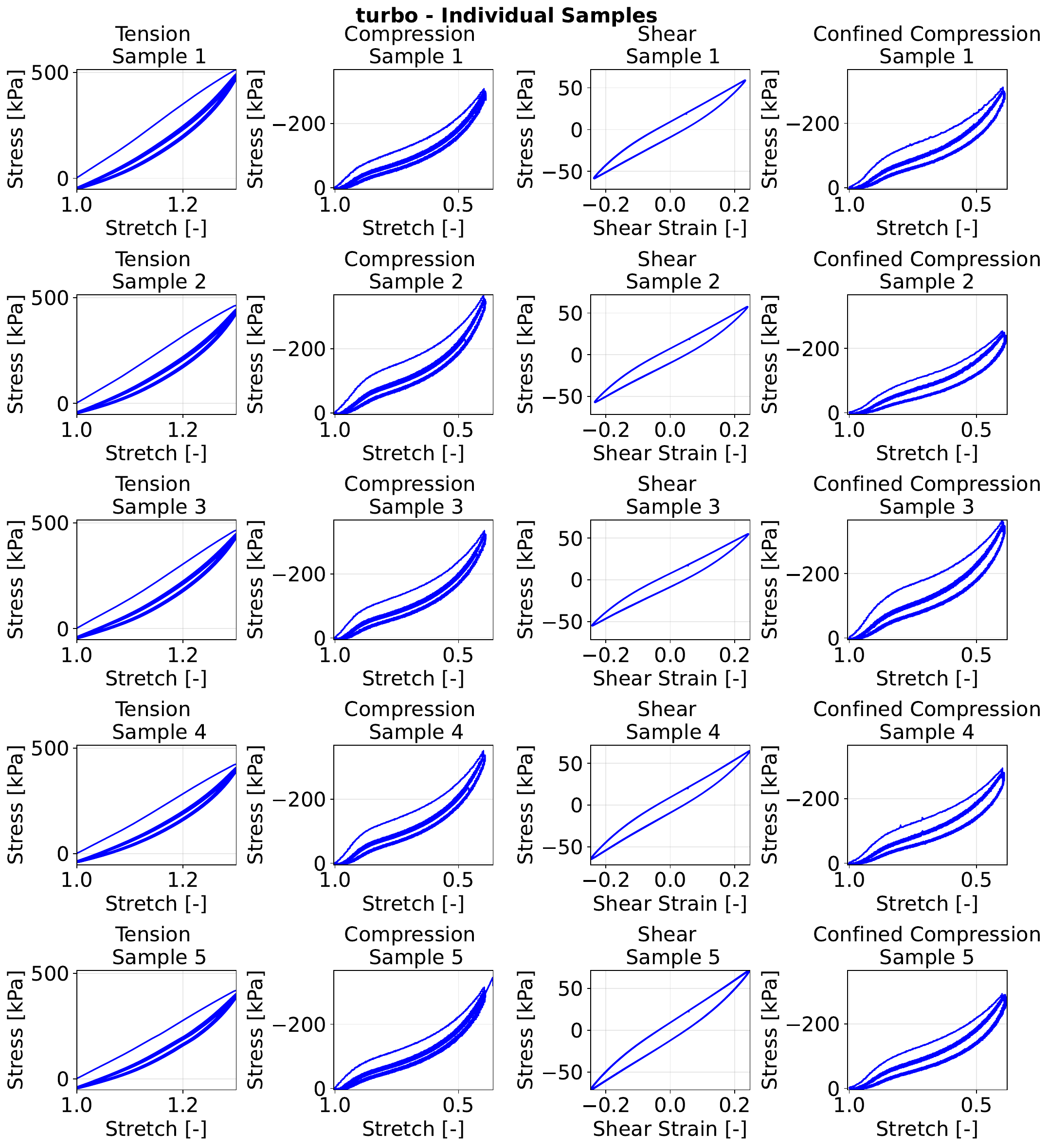}
    \caption{{\sffamily{\bfseries{\turbo raw data from tension, compression, shear, confined compression experiments.}}}
Piola stress vs stretch or shear strain 
measurements for all \turbo samples. 
Columns correspond to the experimental setups, 
tension, compression, shear, and confined compression;
rows corresponds to the different samples.} 
%In order to compute the stretches and stresses from the measured loads and displacements, we used the methods described in section \ref{sec:data_proc} and we used equation \ref{eqn:linear_shear} to compute the shear stress from a single torque measurement. }
    \label{fig:raw_data_turbo}
\end{figure*}

%%%%%%%%%%%%%%%%%%%%%%%%%%%%%%%%%%%%%%%%%%%%%%%%%%%%%%%%%%%%%%%%%%%%%%%%%%%%
Figures \ref{fig:raw_data_leap} and \ref{fig:raw_data_turbo}
summarize the raw data 
from the 
uniaxial tension, 
unconfined compression, 
simple shear, and
confined compression experiments
of the \leap and \turbo foams. 
%%%%%%%%%%%%%%%%%%%%%%%%%%%%%%%%%%%%%%%%%%%%%%%%%%%%%%%%%%%%%%%%%%%%%%%%%%%%
%%%%%%%%%%%%%%%%%%%%%%%%%%%%%%%%%%%%%%%%%%%%%%%%%%%%%%%%%%%%%%%%%%%%%%%%%%%%
%%%%%%%%%%%%%%%%%%%%%%%%%%%%%%%%%%%%%%%%%%%%%%%%%%%%%%%%%%%%%%%%%%%%%%%%%%%%
%%%%%%%%%%%%%%%%%%%%%%%%%%%%%%%%%%%%%%%%%%%%%%%%%%%%%%%%%%%%%%%%%%%%%%%%%%%%
For all experiments and for both foams, 
the shapes of the stress-stretch curves 
appear qualitatively similar across all samples, 
although the magnitude of the stresses 
varies considerably between the different tests.\\[6.pt] 
%%%%%%%%%%%%%%%%%%%%%%%%%%%%%%%%%%%%%%%%%%%%%%%%%%%%%%%%%%%%%%%%%%%%%%%%%%%%
{\sffamily{\bfseries{Uniaxial tension and compression.}}}
%%%%%%%%%%%%%%%%%%%%%%%%%%%%%%%%%%%%%%%%%%%%%%%%%%%%%%%%%%%%%%%%%%%%%%%%%%%%
For the uniaxial tension and compression experiments, 
we see that,
except for the first loading curve,
all loading and unloading curves 
follow similar paths. 
The gap between the unloading and reloading curves is minor, 
which suggests that there is minimal inelastic dissipation; 
however, 
the gap between the first loading curve 
and all subsequent loading curves is clearly noticeble,
which suggests 
a significant Mullins-type stress softening \cite{mullins_softening_1969}, 
an initial conditioning effect with minimal subsequent dissipation.
To disregard this effect,
we exclude the first loading and unloading curves 
from our analysis as described in Section \ref{sec:data_proc}. \\[6.pt]
%%%%%%%%%%%%%%%%%%%%%%%%%%%%%%%%%%%%%%%%%%%%%%%%%%%%%%%%%%%%%%%%%%%%%%%%%%%%
{\sffamily{\bfseries{Simple shear.}}}
%%%%%%%%%%%%%%%%%%%%%%%%%%%%%%%%%%%%%%%%%%%%%%%%%%%%%%%%%%%%%%%%%%%%%%%%%%%%
For the simple shear experiments, 
the gap between the loading and unloading curves is significant, 
which suggests the presence of notable inelastic dissipation. 
The mean of the loading and unloading curves 
is approximately linear, 
which suggests that in the tested range, 
the elastic stress response remains fairly linear.
%%%%%%%%%%%%%%%%%%%%%%%%%%%%%%%%%%%%%%%%%%%%%%%%%%%%%%%%%%%%%%%%%%%%%%%%%%%%
\begin{table*}[h]
\caption{{\sffamily{\bfseries{\leap data from tension, compression, shear experiments.}}} 
Recorded stretch-stress pairs 
$\{ \lambda , P_{11} \}$ and $\{ \gamma, P_{12}\}$, 
linear elastic stiffnesses $\textsf{E}$, and 
relative energy return $\eta$ for the \leap foam.
The first two columns represent uniaxial tension, 
the middle two columns uniaxial compression, and 
the last two columns simple shear. 
Means and standard deviations are reported across $n=5$ samples.} 
\vspace*{0.1cm}
\small
\centering
\label{table:leap}
%%%%%%%%%%%%%%%%%%%%%%%%%%%%%%%%%%%%%%%%%%%%%%%%%%%%%%%%%%%%%%%%%%%%%%%%%%%%
%\resizebox{\textwidth}{!}{%
\begin{tabular}{|cc||cc||cc|}
\hline
  \multicolumn{2}{|c||}{\sffamily{\bfseries{uniaxial tension}}}
& \multicolumn{2}{c||} {\sffamily{\bfseries{uniaxial compression}}}
& \multicolumn{2}{c|}  {\sffamily{\bfseries{simple shear}}} \\
  \multicolumn{2}{|c||}{$n=5$}
& \multicolumn{2}{c||}{$n=5$}
& \multicolumn{2}{c|}{$n=5$} \\ \hline
$\lambda$ & $P_{11}$ & $\lambda$ & $P_{11}$ & $\gamma$ & $P_{12}$  \\
\,[-] & [kPa]  & [-] & [kPa]  & [-] & [kPa]  \\
\hline \hline
1.000 & \phantom{0}\phantom{0}0.00\hspace{0.5em}$\pm$ \phantom{0}\phantom{0}0.00 & 1.000 & \phantom{0}\phantom{0}0.00\hspace{0.5em}$\pm$ \phantom{0}\phantom{0}0.00 & 0.000 & \phantom{0}\phantom{0}0.09\hspace{0.5em}$\pm$ \phantom{0}\phantom{0}0.02
 \\ \hline
1.025 & \phantom{0}13.71\hspace{0.5em}$\pm$ \phantom{0}\phantom{0}2.15 & 0.950 & \phantom{0}12.79\hspace{0.5em}$\pm$ \phantom{0}\phantom{0}1.24 & 0.012 & \phantom{0}\phantom{0}1.42\hspace{0.5em}$\pm$ \phantom{0}\phantom{0}0.28
 \\
1.050 & \phantom{0}29.76\hspace{0.5em}$\pm$ \phantom{0}\phantom{0}4.50 & 0.900 & \phantom{0}32.65\hspace{0.5em}$\pm$ \phantom{0}\phantom{0}3.98 & 0.025 & \phantom{0}\phantom{0}2.85\hspace{0.5em}$\pm$ \phantom{0}\phantom{0}0.56
 \\
1.075 & \phantom{0}46.83\hspace{0.5em}$\pm$ \phantom{0}\phantom{0}7.19 & 0.850 & \phantom{0}45.64\hspace{0.5em}$\pm$ \phantom{0}\phantom{0}5.47 & 0.037 & \phantom{0}\phantom{0}4.29\hspace{0.5em}$\pm$ \phantom{0}\phantom{0}0.85
 \\ \hline
1.100 & \phantom{0}65.35\hspace{0.5em}$\pm$ \phantom{0}10.34 & 0.800 & \phantom{0}55.37\hspace{0.5em}$\pm$ \phantom{0}\phantom{0}5.84 & 0.050 & \phantom{0}\phantom{0}5.75\hspace{0.5em}$\pm$ \phantom{0}\phantom{0}1.14
 \\ \hline
1.125 & \phantom{0}85.64\hspace{0.5em}$\pm$ \phantom{0}13.94 & 0.750 & \phantom{0}65.47\hspace{0.5em}$\pm$ \phantom{0}\phantom{0}6.08 & 0.062 & \phantom{0}\phantom{0}7.24\hspace{0.5em}$\pm$ \phantom{0}\phantom{0}1.45
 \\
1.150 & 107.77\hspace{0.5em}$\pm$ \phantom{0}17.66 & 0.700 & \phantom{0}77.17\hspace{0.5em}$\pm$ \phantom{0}\phantom{0}6.33 & 0.075 & \phantom{0}\phantom{0}8.76\hspace{0.5em}$\pm$ \phantom{0}\phantom{0}1.77
 \\
1.175 & 131.58\hspace{0.5em}$\pm$ \phantom{0}21.25 & 0.650 & \phantom{0}91.21\hspace{0.5em}$\pm$ \phantom{0}\phantom{0}6.69 & 0.087 & \phantom{0}10.34\hspace{0.5em}$\pm$ \phantom{0}\phantom{0}2.11
 \\ \hline
1.200 & 157.34\hspace{0.5em}$\pm$ \phantom{0}24.86 & 0.600 & 108.06\hspace{0.5em}$\pm$ \phantom{0}\phantom{0}6.98 & 0.100 & \phantom{0}11.97\hspace{0.5em}$\pm$ \phantom{0}\phantom{0}2.46
 \\ \hline
1.225 & 185.70\hspace{0.5em}$\pm$ \phantom{0}28.77 & 0.550 & 129.08\hspace{0.5em}$\pm$ \phantom{0}\phantom{0}7.59 & 0.112 & \phantom{0}13.67\hspace{0.5em}$\pm$ \phantom{0}\phantom{0}2.84
 \\
1.250 & 217.85\hspace{0.5em}$\pm$ \phantom{0}33.26 & 0.500 & 156.19\hspace{0.5em}$\pm$ \phantom{0}\phantom{0}8.50 & 0.125 & \phantom{0}15.43\hspace{0.5em}$\pm$ \phantom{0}\phantom{0}3.23
 \\
1.275 & 255.28\hspace{0.5em}$\pm$ \phantom{0}38.67 & 0.450 & 191.67\hspace{0.5em}$\pm$ \phantom{0}10.09 & 0.137 & \phantom{0}17.29\hspace{0.5em}$\pm$ \phantom{0}\phantom{0}3.67
 \\ \hline
1.300 & 298.59\hspace{0.5em}$\pm$ \phantom{0}45.29 & 0.400 & 241.60\hspace{0.5em}$\pm$ \phantom{0}13.73 & 0.150 & \phantom{0}19.14\hspace{0.5em}$\pm$ \phantom{0}\phantom{0}4.12
 \\
\hline \hline
  \multicolumn{2}{|c||}{\sffamily{\bfseries{tensile stiffness}}}
& \multicolumn{2}{c||} {\sffamily{\bfseries{compressive stiffness}}}
& \multicolumn{2}{c|}  {\sffamily{\bfseries{shear stiffness}}} \\
  \multicolumn{2}{|c||}{$\textsf{E}_{\rm{ten}} = 623.65 \pm 96.36$\,kPa}
& \multicolumn{2}{c||} {$\textsf{E}_{\rm{com}} = 299.22 \pm 29.09$\,kPa}
& \multicolumn{2}{c|}  {$\textsf{G}_{\rm{shr}} = 117.16 \pm 23.73$\,kPa} \\
\hline \hline
  \multicolumn{2}{|c||}{\sffamily{\bfseries{relative energy return}}}
& \multicolumn{2}{c||} {\sffamily{\bfseries{relative energy return}}}
& \multicolumn{2}{c|}  {\sffamily{\bfseries{relative energy return}}} \\
  \multicolumn{2}{|c||}{$\eta_{\rm{ten}}  = 90.7 \pm 1.1 \%$}
& \multicolumn{2}{c||} {$\eta_{\rm{com}}  = 89.5 \pm 1.6\%$}
& \multicolumn{2}{c|}  {$\eta_{\rm{shr}}  = 73.6 \pm 0.7 \%$} \\
\hline
\end{tabular}%% End leap table
%\end{table*}
%%%%%%%%%%%%%%%%%%%%%%%%%%%%%%%%%%%%%%%%%%%%%%%%%%%%%%%%%%%%%%%%%%%%%%%%%%%%
\vspace*{0.4cm}
%%%%%%%%%%%%%%%%%%%%%%%%%%%%%%%%%%%%%%%%%%%%%%%%%%%%%%%%%%%%%%%%%%%%%%%%%%%%
%\begin{table*}[h]
\caption{{\sffamily{\bfseries{\turbo data from tension, compression, shear experiments.}}} 
Recorded stretch-stress pairs 
$\{ \lambda , P_{11} \}$ and $\{ \gamma, P_{12}\}$, 
linear elastic stiffnesses $\textsf{E}$, and 
relative energy return $\eta$ for the \turbo foam.
The first two columns represent uniaxial tension, 
the middle two columns uniaxial compression, and 
the last two columns simple shear. 
Means and standard deviations are reported across $n=5$ samples.} 
\vspace*{0.1cm}
\small
\centering
\label{table:turbo}
\begin{tabular}{|cc||cc||cc|}
\hline
  \multicolumn{2}{|c||}{\sffamily{\bfseries{uniaxial tension}}}
& \multicolumn{2}{c||} {\sffamily{\bfseries{uniaxial compression}}}
& \multicolumn{2}{c|}  {\sffamily{\bfseries{simple shear}}} \\
  \multicolumn{2}{|c||}{$n=5$}
& \multicolumn{2}{c||}{$n=5$}
& \multicolumn{2}{c|}{$n=5$} \\ \hline
$\lambda$ & $P_{11}$ & $\lambda$ & $-P_{11}$ & $\gamma$ & $P_{12}$  \\
\,[-] & [kPa]  & [-] & [kPa]  & [-] & [kPa]  \\
\hline \hline
1.000 & \phantom{0}\phantom{0}0.00\hspace{0.5em}$\pm$ \phantom{0}\phantom{0}0.00 & 1.000 & \phantom{0}\phantom{0}0.00\hspace{0.5em}$\pm$ \phantom{0}\phantom{0}0.00 & 0.000 & \phantom{0}\phantom{0}0.16\hspace{0.5em}$\pm$ \phantom{0}\phantom{0}0.02
 \\ \hline
1.025 & \phantom{0}19.45\hspace{0.5em}$\pm$ \phantom{0}\phantom{0}1.64 & 0.950 & \phantom{0}\phantom{0}9.41\hspace{0.5em}$\pm$ \phantom{0}\phantom{0}1.08 & 0.012 & \phantom{0}\phantom{0}2.66\hspace{0.5em}$\pm$ \phantom{0}\phantom{0}0.26
 \\
1.050 & \phantom{0}41.89\hspace{0.5em}$\pm$ \phantom{0}\phantom{0}3.31 & 0.900 & \phantom{0}32.75\hspace{0.5em}$\pm$ \phantom{0}\phantom{0}1.96 & 0.025 & \phantom{0}\phantom{0}5.33\hspace{0.5em}$\pm$ \phantom{0}\phantom{0}0.52
 \\
1.075 & \phantom{0}66.29\hspace{0.5em}$\pm$ \phantom{0}\phantom{0}5.12 & 0.850 & \phantom{0}51.40\hspace{0.5em}$\pm$ \phantom{0}\phantom{0}4.41 & 0.037 & \phantom{0}\phantom{0}8.03\hspace{0.5em}$\pm$ \phantom{0}\phantom{0}0.77
 \\ \hline
1.100 & \phantom{0}93.27\hspace{0.5em}$\pm$ \phantom{0}\phantom{0}7.29 & 0.800 & \phantom{0}64.33\hspace{0.5em}$\pm$ \phantom{0}\phantom{0}5.62 & 0.050 & \phantom{0}10.76\hspace{0.5em}$\pm$ \phantom{0}\phantom{0}1.03
 \\ \hline
1.125 & 123.76\hspace{0.5em}$\pm$ \phantom{0}\phantom{0}9.81 & 0.750 & \phantom{0}76.82\hspace{0.5em}$\pm$ \phantom{0}\phantom{0}6.22 & 0.062 & \phantom{0}13.54\hspace{0.5em}$\pm$ \phantom{0}\phantom{0}1.28
 \\
1.150 & 157.97\hspace{0.5em}$\pm$ \phantom{0}12.40 & 0.700 & \phantom{0}90.92\hspace{0.5em}$\pm$ \phantom{0}\phantom{0}6.66 & 0.075 & \phantom{0}16.39\hspace{0.5em}$\pm$ \phantom{0}\phantom{0}1.53
 \\
1.175 & 195.56\hspace{0.5em}$\pm$ \phantom{0}14.99 & 0.650 & 107.86\hspace{0.5em}$\pm$ \phantom{0}\phantom{0}7.34 & 0.087 & \phantom{0}19.33\hspace{0.5em}$\pm$ \phantom{0}\phantom{0}1.77
 \\ \hline
1.200 & 236.36\hspace{0.5em}$\pm$ \phantom{0}18.01 & 0.600 & 128.72\hspace{0.5em}$\pm$ \phantom{0}\phantom{0}8.32 & 0.100 & \phantom{0}22.35\hspace{0.5em}$\pm$ \phantom{0}\phantom{0}2.02
 \\ \hline
1.225 & 281.58\hspace{0.5em}$\pm$ \phantom{0}21.73 & 0.550 & 155.11\hspace{0.5em}$\pm$ \phantom{0}\phantom{0}9.66 & 0.112 & \phantom{0}25.47\hspace{0.5em}$\pm$ \phantom{0}\phantom{0}2.27
 \\
1.250 & 333.96\hspace{0.5em}$\pm$ \phantom{0}25.42 & 0.500 & 189.78\hspace{0.5em}$\pm$ \phantom{0}11.63 & 0.125 & \phantom{0}28.69\hspace{0.5em}$\pm$ \phantom{0}\phantom{0}2.50
 \\
1.275 & 395.95\hspace{0.5em}$\pm$ \phantom{0}29.71 & 0.450 & 236.96\hspace{0.5em}$\pm$ \phantom{0}14.90 & 0.137 & \phantom{0}32.03\hspace{0.5em}$\pm$ \phantom{0}\phantom{0}2.76
 \\ \hline
1.300 & 469.95\hspace{0.5em}$\pm$ \phantom{0}34.89 & 0.400 & 305.43\hspace{0.5em}$\pm$ \phantom{0}19.56 & 0.150 & \phantom{0}35.41\hspace{0.5em}$\pm$ \phantom{0}\phantom{0}2.99
 \\
\hline \hline
  \multicolumn{2}{|c||}{\sffamily{\bfseries{tensile stiffness}}}
& \multicolumn{2}{c||} {\sffamily{\bfseries{compressive stiffness}}}
& \multicolumn{2}{c|}  {\sffamily{\bfseries{shear stiffness}}} \\
  \multicolumn{2}{|c||}{$\textsf{E}_{\rm{ten}} = 884.15 \pm 68.81$\,kPa}
& \multicolumn{2}{c||} {$\textsf{E}_{\rm{com}} = 267.94 \pm 15.67$\,kPa}
& \multicolumn{2}{c|}  {$\textsf{G}_{\rm{shr}} = 219.12 \pm 20.39$\,kPa} \\
\hline \hline
  \multicolumn{2}{|c||}{\sffamily{\bfseries{relative energy return}}}
& \multicolumn{2}{c||} {\sffamily{\bfseries{relative energy return}}}
& \multicolumn{2}{c|}  {\sffamily{\bfseries{relative energy return}}} \\
  \multicolumn{2}{|c||}{$\eta_{\rm{ten}}  = 94.3 \pm 1.3 \%$}
& \multicolumn{2}{c||} {$\eta_{\rm{com}}  = 84.6 \pm 1.3\%$}
& \multicolumn{2}{c|}  {$\eta_{\rm{shr}}  = 75.6 \pm 0.7 \%$} \\
\hline
\end{tabular}
\end{table*}
%%%%%%%%%%%%%%%%%%%%%%%%%%%%%%%%%%%%%%%%%%%%%%%%%%%%%%%%%%%%%%%%%%%%%%%%%%%%
%%%%%%%%%%%%%%%%%%%%%%%%%%%%%%%%%%%%%%%%%%%%%%%%%%%%%%%%%%%%%%%%%%%%%%%%%%%%
\begin{figure*}
%\centering
\hspace*{1.2cm}
\includegraphics[width=0.42\linewidth]{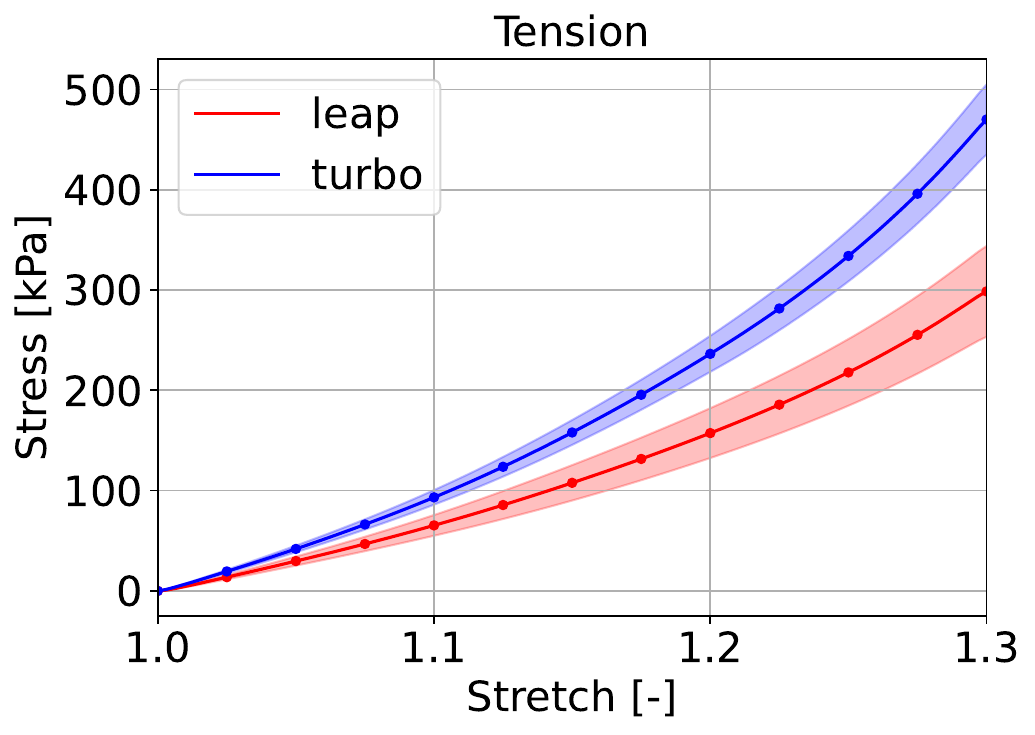}\includegraphics[width=0.42\linewidth]{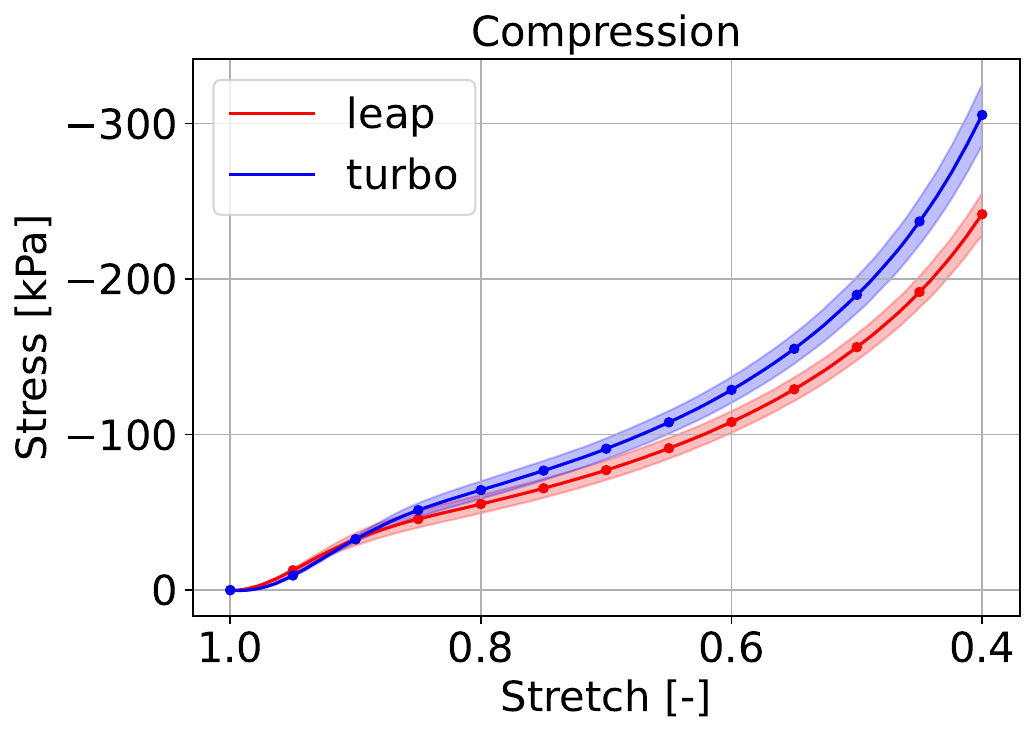} 
\newline
\hspace*{1.2cm}
\includegraphics[width=0.42\linewidth]{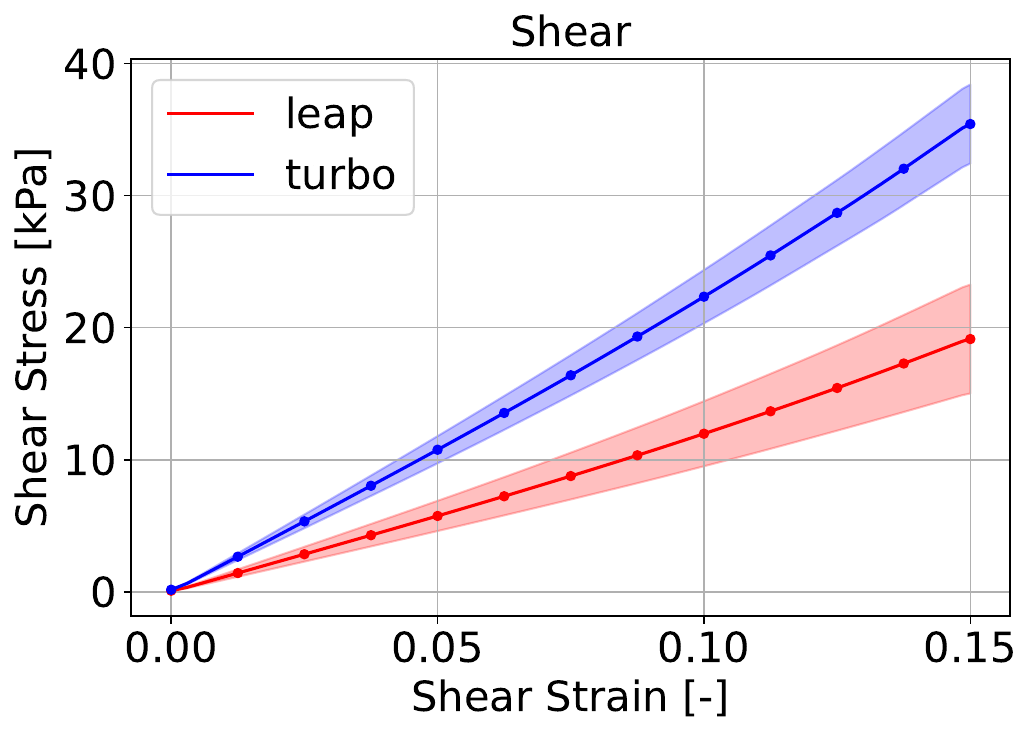}
\includegraphics[width=0.42\linewidth]{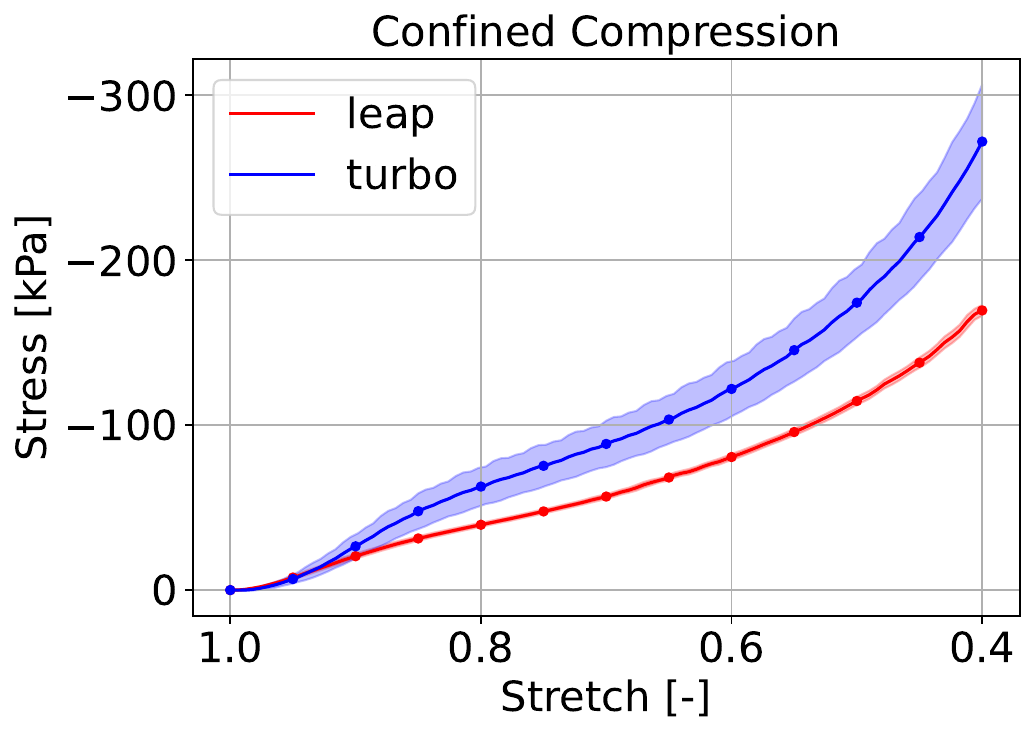}
\caption{{\sffamily{\bfseries{\leap and \turbo data from tension, compression, shear, and confined compression experiments.}}} 
Recorded stretch-stress curves 
$\{ \lambda , P_{11} \}$ and $\{ \gamma, P_{12}\}$, 
for the \leap  foam, shown in red, and 
the \turbo foam, shown in blue. 
Dots highlight the discrete values from Tables \ref{table:leap} and \ref{table:turbo}, 
lines represent the means, and 
shaded areas the standard deviations across $n=5$ samples.}
%Means and standard deviations are reported across $n=5$ samples.}
%These figures show the averaged stress values for all four experiments. The shaded area shows one standard deviation above and below the mean. }
    \label{fig:proc_data_all}
\end{figure*}
%%%%%%%%%%%%%%%%%%%%%%%%%%%%%%%%%%%%%%%%%%%%%%%%%%%%%%%%%%%%%%%%%%%%%%%%%%%%
\\[6.pt]
%%%%%%%%%%%%%%%%%%%%%%%%%%%%%%%%%%%%%%%%%%%%%%%%%%%%%%%%%%%%%%%%%%%%%%%%%%%%
{\sffamily{\bfseries{Unconfined and confined compression.}}}
%%%%%%%%%%%%%%%%%%%%%%%%%%%%%%%%%%%%%%%%%%%%%%%%%%%%%%%%%%%%%%%%%%%%%%%%%%%%
For the confined compression experiments, 
the stress is comparable to the unconfined compression experiment, 
but slightly lower. 
We cannot explain this observation 
with a standard constitutive model, 
as constraining the transverse stretch 
should only increase the strain energy 
compared to unconfined compression.
The observed differences  
likely reflect experimental artifacts, 
frictional effects, imperfect confinement, 
or compliance of the mounting system, 
rather than an intrinsic material response.
We conclude 
that the transverse stretch 
in the unconfined and confined compression experiments 
is similar and
equal to, or at least very close to one.
Taken together, 
the similarity of the confined and unconfined compression experiments 
is consistent with a small effective Poisson's ratio in compression.
%%%%%%%%%%%%%%%%%%%%%%%%%%%%%%%%%%%%%%%%%%%%%%%%%%%%%%%%%%%%%%%%%%%%%%%%%%%
\begin{table*}[h]
\caption{{\sffamily{\bfseries{Performance of discovered models for \leap and \turbo.}}} 
Activated terms, regularization parameter $\alpha$, 
number of non-zero terms, and
goodness of fit ${\textsf{R}}^2$ in tension, compression, and shear. 
In each row, we activate a subset of the terms of the network in Figure \ref{fig:cann-arch}, with 
SI single-invariant, 
MI mixed-invariant terms, and 
PS principal-stretch terms. 
We apply either no regularization, with $\alpha =0.0$,
or $\textsf{L}_{0.5}$ regularization with $\alpha=1.0$. 
The gray rows highlight the models in 
Figures \ref{fig:si_mi} and \ref{fig:si_ps}.} 
\vspace*{0.2cm}
\small
\centering
\label{table:model_r2}
%%%%%%%%%%%%%%%%%%%%%%%%%%%%%%%%%%%%%%%%%%%%%%%%%%%%%%%%%%%%%%%%%%%%%%%%%%%%
%\resizebox{\textwidth}{!}{
\begin{tabular}{|c|c|c||c|ccc||c|ccc|}
\hline
%\multicolumn{1}{|c}
\multicolumn{3}{|c||}{{\sffamily{\bfseries{\; \;\, network}}}}& 
%\multicolumn{2}{|c||}{{\sffamily{\bfseries{network}}}}& 
\multicolumn{4}{c||}{\leap}& 
\multicolumn{4}{c|}{\turbo}\\ \hline
%& 
& \makecell{activated \\ terms} 
& $\alpha$ 
&   \makecell{\!\!non-zero\!\! \\ terms}  
& ${\textsf{R}}_{\rm{ten}}^2$ 
& ${\textsf{R}}_{\rm{com}}^2$ 
& ${\textsf{R}}_{\rm{shr}}^2$ 
&   \makecell{\!\!non-zero\!\! \\ terms} 
& ${\textsf{R}}_{\rm{ten}}^2$ 
& ${\textsf{R}}_{\rm{com}}^2$ 
& ${\textsf{R}}_{\rm{shr}}^2$ \\ \hline  \hline
  {\textsf{SI}} &
 $\psi_{\bar{I}_1},\psi_{\bar{I}_2},\psi_{J}$
& 0.0 & 6 & 0.000 & 0.891 & 0.983 & 8 & 0.241 & 0.577 & 0.839 \\
 {\textsf{SI}} &
$\psi_{\bar{I}_1},\psi_{\bar{I}_2},\psi_{J}$ 
& 1.0 & 2 & 0.209 & 0.927 & 0.988 & 2 & 0.195 & 0.765 & 0.772 \\  \hline
 {\textsf{SI+MI}} &
 $\!\!\psi_{\bar{I}_1},\psi_{\bar{I}_2},\psi_{J},\psi_{\bar{I}_1,J},\psi_{\bar{I}_2,J}\!\!$ 
& 0.0 & 7 & 0.991 & 0.984 & 0.995 & 7 & 0.999 & 0.979 & 0.996 \\
\rowcolor{black!20}
 {\textsf{SI+MI}} &
 $\!\!\psi_{\bar{I}_1},\psi_{\bar{I}_2},\psi_{J},\psi_{\bar{I}_1,J},\psi_{\bar{I}_2,J}\!\!$  & 1.0 & 3 & 0.981 & 0.988 & 0.995 & 3 & 0.997 & 0.991 & 0.996 \\  \hline
 {\textsf{SI+PS}} &
 $\psi_{\bar{I}_1},\psi_{\bar{I}_2},\psi_{J},\psi_{\lambda_i}$ 
& 0.0 & 6 & 0.989 & 0.997 & 0.999 & 7 & 0.998 & 0.939 & 0.985 \\
\rowcolor{black!20}%{yellow!20}{blue!50!cyan!15}
 {\textsf{SI+PS}} &
$\psi_{\bar{I}_1},\psi_{\bar{I}_2},\psi_{J},\psi_{\lambda_i}$  & 1.0 & 3 & 0.992 & 0.995 & 0.999 & 3 & 0.999 & 0.938 & 0.984 \\
\hline
\end{tabular} %}
\end{table*}%\\[6.pt]
%%%%%%%%%%%%%%%%%%%%%%%%%%%%%%%%%%%%%%%%%%%%%%%%%%%%%%%%%%%%%%%%%%%%%%%%%%%%
%%%%%%%%%%%%%%%%%%%%%%%%%%%%%%%%%%%%%%%%%%%%%%%%%%%%%%%%%%%%%%%%%%%%%%%%%%%%
\subsection{{\sffamily{\bfseries{Data analysis}}}}
%%%%%%%%%%%%%%%%%%%%%%%%%%%%%%%%%%%%%%%%%%%%%%%%%%%%%%%%%%%%%%%%%%%%%%%%%%%%
Tables \ref{table:leap} and \ref{table:turbo}
summarize  
the stretch-stress pairs,
the linear elastic stiffness, and
the relative energy return 
for both foams
in tension, compression, and shear,
averaged across all $n=5$ tests.
Figure~\ref{fig:proc_data_all} 
illustrates the results for
both foams in a side-by-side comparison. \\[6.pt] 
%%%%%%%%%%%%%%%%%%%%%%%%%%%%%%%%%%%%%%%%%%%%%%%%%%%%%%%%%%%%%%%%%%%%%%%%%%%%
{\sffamily{\bfseries{Tension-compression asymmetry.}}}
%%%%%%%%%%%%%%%%%%%%%%%%%%%%%%%%%%%%%%%%%%%%%%%%%%%%%%%%%%%%%%%%%%%%%%%%%%%%
For the \leap, 
the linear tensile, compressive, and shear stiffnesses of 
${\textsf{E}}_{\rm{ten}}=623\pm96$\,kPa and
${\textsf{E}}_{\rm{com}}=299\pm29$\,kPa and
${\textsf{G}}_{\rm{shr}}=117\pm24$\,kPa
clearly highlight a strong tension-compression asymmetry, 
with a more than twice-as-high resistance to tension compared to compression,
and a low resistance to shear, 
consistent with low-density elastomeric foams.
For the \turbo, 
the linear tensile, compressive, and shear stiffnesses of
${\textsf{E}}_{\rm{ten}}=884\pm69$\,kPa and
${\textsf{E}}_{\rm{com}}=268\pm16$\,kPa and
${\textsf{G}}_{\rm{shr}}=219\pm20$\,kPa 
show an even more pronounced tension-compression asymmetry, 
but display a markably higher resistance to shear. 
The relative energy return mimics these trends:
It is compatible for the \leap and \turbo,
highest in tension with
90.7$\pm$1.1\% and 94.3$\pm$1.3\%,
followed by compression with
89.5$\pm$1.6\% and 84.6$\pm$1.3\%, and
lowest in shear with
73.6$\pm$0.9\% and 75.6$\pm$0.7\%.\\[6.pt]
%%%%%%%%%%%%%%%%%%%%%%%%%%%%%%%%%%%%%%%%%%%%%%%%%%%%%%%%%%%%%%%%%%%%%%%%%%%%
{\sffamily{\bfseries{\turbo is stiffer than \leap.}}}
%%%%%%%%%%%%%%%%%%%%%%%%%%%%%%%%%%%%%%%%%%%%%%%%%%%%%%%%%%%%%%%%%%%%%%%%%%%% 
Compared with the \leap, 
in the linear regime,
the \turbo exhibits 
a similar compressive stiffness of -10.4\%
but a significantly larger ($p=0.0017$) tensile stiffness of +41.9\%,
and a substantially larger ($p=0.00007$) shear stiffness of +87.2\%, 
which indicates a markedly greater resistance 
to lateral deformation % or shear 
at a comparable level of vertical support.
This trend continues into the non-linear regime.
At our maximum relative deformations
of 30\% in tension, 60\% in compression, and 15\% in shear,
\turbo exhibits significantly higher stresses than \leap 
in tension ($p = 0.033$), 
compression ($p = 0.001$), and 
shear ($p = 0.0003$), 
with the largest difference in shear:
Its tensile and compressive stresses 
are about 25\% larger, 
while its shear stresses
are almost twice as large. 
%Relative to the mean,
%the standard deviations of the data 
%are smallest in compression, 
%with 6.3\% and 7.2\% for the \leap and \turbo, 
%followed by tension,
%with 19.4\% and 8.7\%, 
%and then shear,
%with 24.1\% and 9.5\%,
%suggesting that 
%the measurements are robust and repeatable overall.
%%%%%%%%%%%%%%%%%%%%%%%%%%%%%%%%%%%%%%%%%%%%%%%%%%%%%%%%%%%%%%%%%%%%%%%%%%%%
\subsection{{\sffamily{\bfseries{Automated model discovery}}}}
%\subsection{Model Training}
%%%%%%%%%%%%%%%%%%%%%%%%%%%%%%%%%%%%%%%%%%%%%%%%%%%%%%%%%%%%%%%%%%%%%%%%%%%%
Next, we discover models for both foams
using the constitutive neural network from Figure \ref{fig:cann-arch}. 
We train on tension, compression, and shear data 
for 15,000 epochs 
with a batch size of 64 and 
a learning rate of 0.01;
where, 
for the first 5,000 epochs, 
we set the regularization parameter $\alpha$ to zero. 
We the vary the network architecture 
to include all ten single-invariant terms 
and either no other terms, or
the two mixed-invariant terms, or 
the two principal-stretch terms. 
For each architecture, we set
the regularization parameter $\alpha$ 
to either 0.0 or 1.0, 
as our preliminary parameter study has shown 
that this reduces the number of terms 
without sacrificing the goodness of fit.
Table \ref{table:model_r2} summarizes the results 
from training the network for all six cases.\\[6.pt]
%%%%%%%%%%%%%%%%%%%%%%%%%%%%%%%%%%%%%%%%%%%%%%%%%%%%%%%%%%%%%%%%%%%%%%%%%%%%
{\sffamily{\bfseries{Performance of discovered models for \leap and \turbo.}}}
%%%%%%%%%%%%%%%%%%%%%%%%%%%%%%%%%%%%%%%%%%%%%%%%%%%%%%%%%%%%%%%%%%%%%%%%%%%%
The network performs poorly 
when using only the single invariant terms, 
especially in tension. 
However, 
when adding either the mixed-invariant terms 
or the principal-stretch terms, 
the goodness of fit improves drastically.
Table \ref{table:model_r2} confirms that
adding regularization 
to either of these two models
significantly reduces the number of terms,
from six or seven down to three,
and has a minimal impact on the overall goodness of fit. 
We focus on these two models
with a regularization parameter of $\alpha = 1.0$,
which provides an excellent goodness of fit
while remaining sparse and interpretable.  
The discovered single-invariant-mixed-invariant
model for both foams is
The discovered single-invariant-mixed-invariant
model for both foams is
%%%%%%%%%%%%%%%%%%%%%%%%%%%%%%%%%%%%%%%%%%%%%%%%%%%%%%%%%%%%%%%%%%%%%%%%%%%%
\[
\begin{array}{l@{\hspace*{0.05cm}}c@{\hspace*{0.1cm}}
              r@{\hspace*{0.1cm}}l}
    \psi^{\rm{leap}} 
&=&  26.9 \mbox{kPa} & \hspace*{0.9cm}[\bar I_1 - 3] \\
&+&  70.3 \mbox{kPa} &\; [J^{1.06} - 1.06 \ln(J) - 1] \\
&+&  79.0 \mbox{kPa} &  \; J^{4.86} \, [\bar I_1 - 3]  \\[6.pt]
     \psi^{\rm{turbo}} 
&=&  936 \mbox{kPa} & \;[\exp(0.0615 \ln(J)^2) - 1] \\
&+& 139 \mbox{kPa} &  \; J^{4.36} \, [\bar I_1 - 3] \\  
&+&  19.8 \mbox{kPa} &  \; J^{1.47} \, [\bar I_2 - 3] \,,

\end{array}
\]
%%%%%%%%%%%%%%%%%%%%%%%%%%%%%%%%%%%%%%%%%%%%%%%%%%%%%%%%%%%%%%%%%%%%%%%%%%%%
and the discovered single-invariant-principal-stretch
model for both foam is
%%%%%%%%%%%%%%%%%%%%%%%%%%%%%%%%%%%%%%%%%%%%%%%%%%%%%%%%%%%%%%%%%%%%%%%%%%%%
\[
\begin{array}{l@{\hspace*{0.05cm}}c@{\hspace*{0.1cm}}
              r@{\hspace*{0.1cm}}l}
    \psi^{\rm{leap}}  
&=&  9.93 \mbox{kPa} & [\bar I_1 - 3]^2 \\
&+& 59.0 \mbox{kPa} & [J^{0.481} - 0.481 \ln(J) - 1] \\
&+& 0.286 \mbox{kPa} & \sum_{i=1}^3 
                      [\lambda_i^{8.40} - 8.40 \ln(\lambda_i) - 1] 
                      
                      \\[6.pt]
    \psi^{\rm{turbo}} 
&=& 73.9    \mbox{kPa} &[\,\exp(0.147 (\bar I_1 - 3)) - 1] \\
&+& 0.00592 \mbox{kPa} &[\,\exp(6.64 \ln(J)^2) - 1] \\
&+& 0.365   \mbox{kPa} & \sum_{i=1}^3 
                        [\lambda_i^{8.33} - 8.33 \ln(\lambda_i) - 1] \,.
\end{array}
\]
%%%%%%%%%%%%%%%%%%%%%%%%%%%%%%%%%%%%%%%%%%%%%%%%%%%%%%%%%%%%%%%%%%%%%%%%%%%%
%%%%%%%%%%%%%%%%%%%%%%%%%%%%%%%%%%%%%%%%%%%%%%%%%%%%%%%%%%%%%%%%%%%%%%%%%%%%
\begin{figure*}
    \centering
    \includegraphics[width=1.0\linewidth]{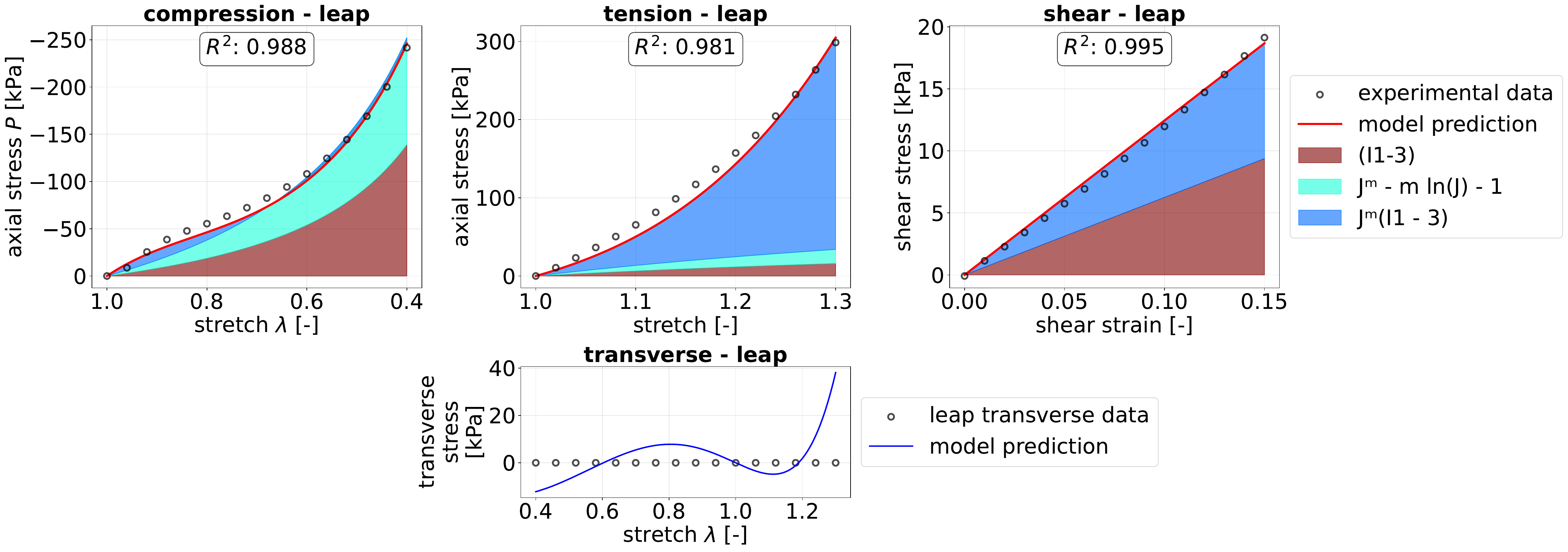}
    \includegraphics[width=1.0\linewidth]{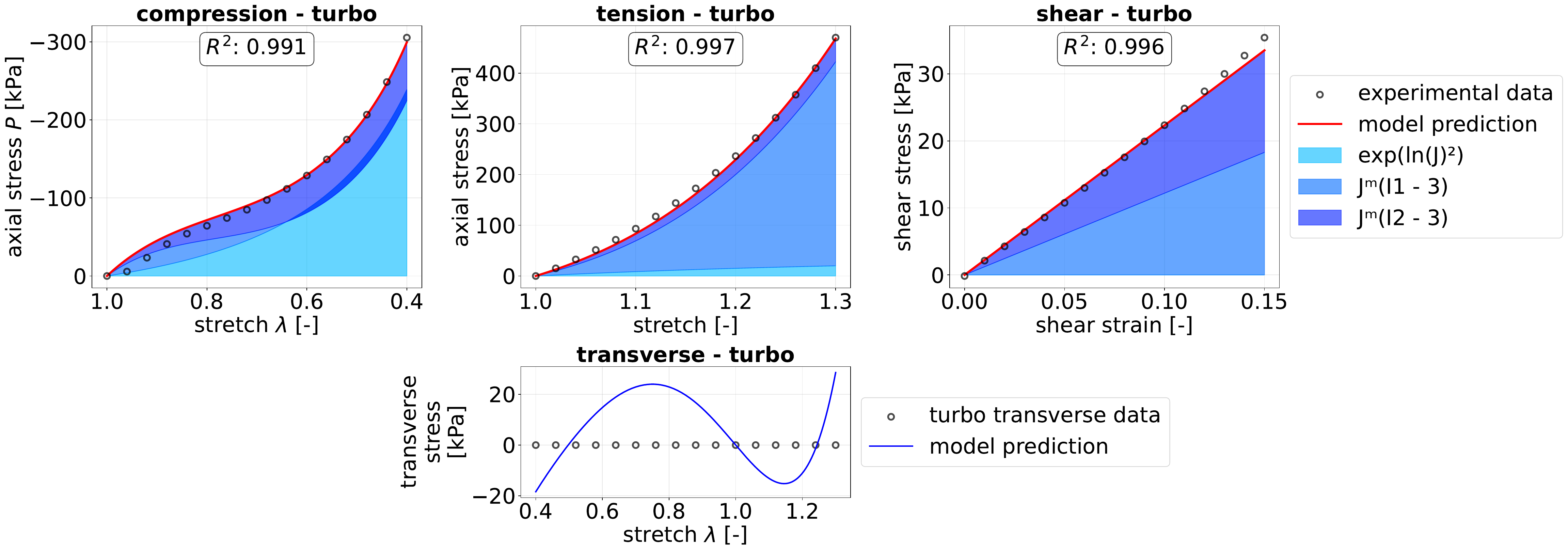}
%%%%%%%%%%%%%%%%%%%%%%%%%%%%%%%%%%%%%%%%%%%%%%%%%%%%%%%%%%%%%%%%%%%%%%%%%%%%
    \caption{{\sffamily{\bfseries{Discovered model with single-invariant and mixed-invariant terms for \leap and \turbo.}}} 
Model prediction of constitutive neural network trained with the single-invariant and mixed-invariant terms and $\textsf{L}_{0.5}$ regularization with a  regularization parameter $\alpha = 1.0$. For tension, compression, and shear, the contributions from each term are shown in a different color and the ${\textsf{R}}^2$ value between model and data is labeled in each plot. Note that the mixed invariant terms may have a negative stress contribution in compression.}
    \label{fig:si_mi}
\end{figure*}
%%%%%%%%%%%%%%%%%%%%%%%%%%%%%%%%%%%%%%%%%%%%%%%%%%%%%%%%%%%%%%%%%%%%%%%%%%%%
%%%%%%%%%%%%%%%%%%%%%%%%%%%%%%%%%%%%%%%%%%%%%%%%%%%%%%%%%%%%%%%%%%%%%%%%%%%%
\begin{figure*}
    \centering
    \includegraphics[width=1.0\linewidth]{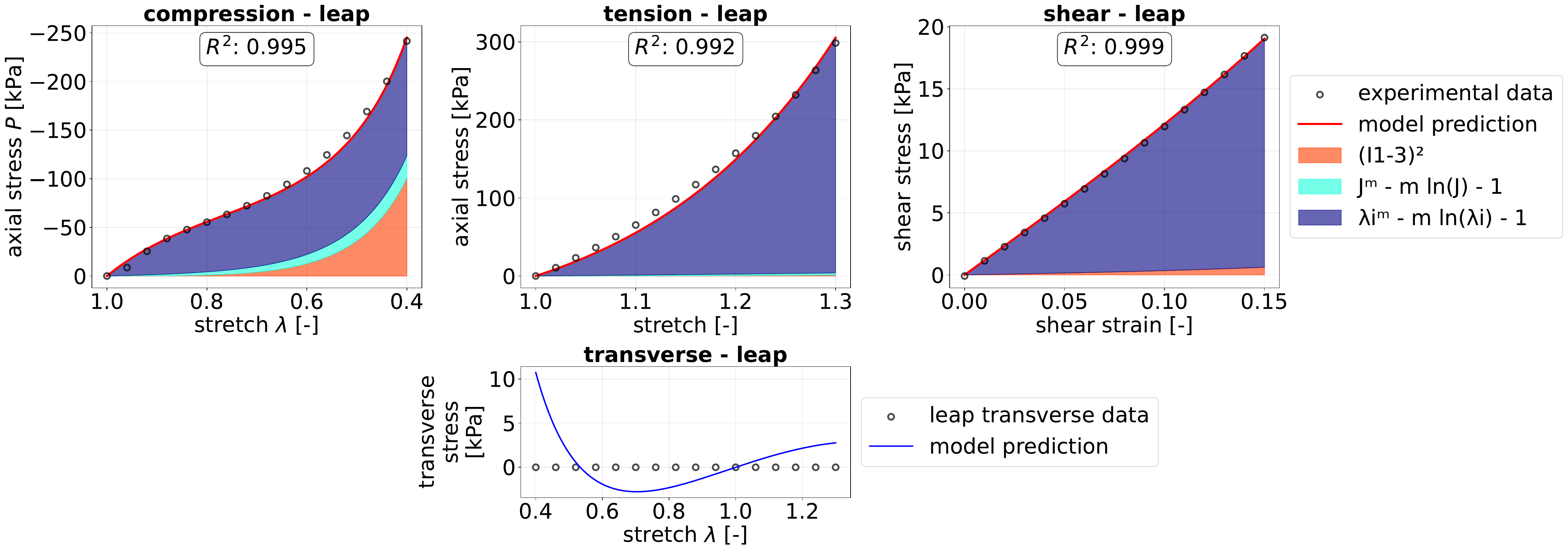}
    \includegraphics[width=1.0\linewidth]{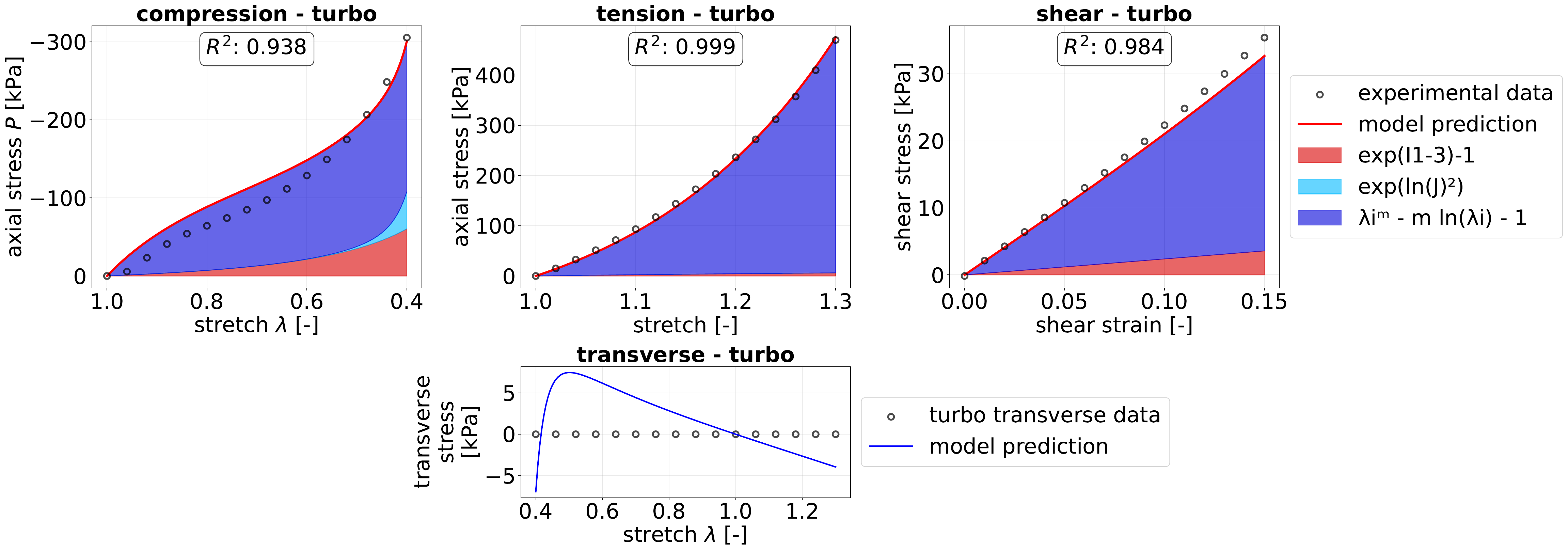}
%%%%%%%%%%%%%%%%%%%%%%%%%%%%%%%%%%%%%%%%%%%%%%%%%%%%%%%%%%%%%%%%%%%%%%%%%%%%
    \caption{{\sffamily{\bfseries{Discovered model with single-invariant and principal-stretch terms for \leap and \turbo.}}} 
Model prediction of constitutive neural network trained with the single-invariant and principal-stretch terms and $\textsf{L}_{0.5}$ regularization with a  regularization parameter $\alpha = 1.0$. For tension, compression, and shear, the contributions from each term are shown in a different color and the ${\textsf{R}}^2$ value between model and data is labeled in each plot.}    \label{fig:si_ps}
\end{figure*}
%%%%%%%%%%%%%%%%%%%%%%%%%%%%%%%%%%%%%%%%%%%%%%%%%%%%%%%%%%%%%%%%%%%%%%%%%%%%
Figures \ref{fig:si_mi} and \ref{fig:si_ps}
show the predicted stress for these two models 
with the contributions 
from the three individual terms
highlighted in different colors.
%%%%%%%%%%%%%%%%%%%%%%%%%%%%%%%%%%%%%%%%%%%%%%%%%%%%%%%%%%%%%%%%%%%%%%%%%%%%
\section{\sffamily{\bfseries{Discussion}}}
%%%%%%%%%%%%%%%%%%%%%%%%%%%%%%%%%%%%%%%%%%%%%%%%%%%%%%%%%%%%%%%%%%%%%%%%%%%%
We successfully quantified 
ultra-low-density elastomeric midsole foams in
uniaxial tension, 
unconfined and confined compression, 
and simple shear 
using a consistent experimental protocol 
that resulted in robust and repeatable measurements 
with low variability across samples. 
These data enabled the discovery 
of sparse, interpretable constitutive models 
that capture the essential mechanical characteristics 
of each foam across all loading modes. \\[6.pt]
%%%%%%%%%%%%%%%%%%%%%%%%%%%%%%%%%%%%%%%%%%%%%%%%%%%%%%%%%%%%%%%%%%%%%%%%%%%%
{\sffamily{\bfseries{Ultra-low-density elastomeric foams
display an unconventional, unique mechanical signature.}}}
%%%%%%%%%%%%%%%%%%%%%%%%%%%%%%%%%%%%%%%%%%%%%%%%%%%%%%%%%%%%%%%%%%%%%%%%%%%%
First and foremost, 
our experimental tests of both foams
reveal the characteristic features 
of ultra-low-density elastomeric polymeric foams:
a pronounced tension–compression asymmetry and 
a relatively low shear resistance \cite{gibson_1997}. 
The close agreement 
between the unconfined and confined compression tests 
of both foams
indicates a negligible lateral strain, 
consistent with an effective Poisson's ratio close to zero,
$\nu \approx 0$ \cite{lakes_negative_poisson_1987}.
The narrow loading–unloading loops 
in both tension and compression indicate 
a low hysteresis and 
an efficient energy return \cite{gent_engineering_2012}, 
91\% and 94\% in tension and
90\% and 85\% in compression,
consistent with the high energy return 
from 76\% to 87\%
reported in elite-level racing shoes
\cite{hoogkamer_comparison_2018}.
For both foams, 
tensile stiffness exceeds compressive stiffness, 
consistent with bending-dominated cellular architectures, 
in which compression and shear primarily engage cell wall bending, 
whereas tension activates cell wall stretching \cite{gibson_1997}. 
Both foams 
provide comparable compressive support, 
268\,kPa vs. 299\,kPa, 
indicating similar vertical cushioning 
under body-weight loading. 
In contrast, 
the \turbo exhibits a 42\% higher tensile stiffness,
884\,kPa vs. 624\,kPa,
and nearly double the shear stiffness of the \leap, 
219\,kPa vs. 117\,kPa,
which implies that it provides
substantially greater resistance to lateral deformation. 
These differences are consistent 
with the design philosophy of the Asics Metaspeed series, 
in which the \edge and \sky variants differ 
primarily in the vertical arrangement 
of these two foams relative to the carbon-fiber plate. 
Interestingly,
the newest member of the Asics Metaspeed series, 
the \ray variant,
only uses the \leap, both above and below the carbon-fiber plate. 
As long-distance road running 
is dominated by sagittal-plane loading, 
the low compressive stiffness of both foams 
appears well suited for straight-line running efficiency. 
However,
during sharp cornering or curved-path running,
the lower shear stiffness of the \leap
could increase injury risk, while
the higher shear stiffness of the \turbo
provides enhanced lateral support and improved stability,
especially in curvy and uneven terrain.\\[6.pt]
%The Metaspeed series 
%utilizes Asics Flight Foam Leap technology 
%to enhance speed and responsiveness. 
%%%%%%%%%%%%%%%%%%%%%%%%%%%%%%%%%%%%%%%%%%%%%%%%%%%%%%%%%%%%%%%%%%%%%%%%%%%%
{\sffamily{\bfseries{Single-invariant models perform poorly.}}}
%%%%%%%%%%%%%%%%%%%%%%%%%%%%%%%%%%%%%%%%%%%%%%%%%%%%%%%%%%%%%%%%%%%%%%%%%%%%
For both foams, 
the classical network \cite{linka_new_2023}
with only single invariant terms,
$\psi_{\bar{I}_1},\psi_{\bar{I}_2},\psi_J$, 
fits the data quite poorly, 
especially in tension. 
The $\textsf{R}^2$ values of its
$\textsf{L}_{0.5}$ regularized version with a penalty of $\alpha = 1.0$
range from 
$\textsf{R}_{\rm{ten}}^2=0.20$ in tension to  
$\textsf{R}_{\rm{com}}^2=0.77$ and $\textsf{R}_{\rm{com}}^2=0.93$ in compression, to
$\textsf{R}_{\rm{shr}}^2=0.77$ and $\textsf{R}_{\rm{shr}}^2=0.99$ in shear.
This suggests that, 
without mixed-invariant 
or principal-stretch terms \cite{linka_brain_2023}, 
the network cannot accurately capture 
the characteristic tension-compression asymmetry 
of ultra-low density elastomeric foams. 
\\[6.pt]
%%%%%%%%%%%%%%%%%%%%%%%%%%%%%%%%%%%%%%%%%%%%%%%%%%%%%%%%%%%%%%%%%%%%%%%%%%%%
{\sffamily{\bfseries{Single-invariant-mixed-invariant models perform excellently.}}}
%%%%%%%%%%%%%%%%%%%%%%%%%%%%%%%%%%%%%%%%%%%%%%%%%%%%%%%%%%%%%%%%%%%%%%%%%%%%
For the single-invariant-mixed-invariant model 
with an $\textsf{L}_{0.5}$ regularization penalty of $\alpha = 1.0$, 
the discovered models for both foams 
fit the data accurately with only three terms. 
Their $\textsf{R}^2$ values 
range from 
$\textsf{R}^2=0.99$ to $\textsf{R}^2=1.00$ 
for both foams across all three experiments.
For both foam models,
the single-invariant terms 
account for most of stress in compression, 
while the mixed-invariant terms 
account for more of the stress in shear 
and almost the entire stress in tension. 
The transverse stress remains close to zero 
across the entire stretch regime, 
and does not exceed 40\,kPa in magnitude. 
Thus, 
we conclude that the mixed invariant terms 
are necessary \cite{blatz_ko_rubber_1962}
to achieve a strong tension-compression asymmetry 
with a much greater stiffness in tension. \\[6.pt] 
%\textcolor{red}{The primary downside to these models is that both the $J^m (\bar I_1 - 3)$ term and the $J^m (\bar I_2 - 3)$ term  are not polyconvex, even when summed with other convex terms.} \\[6.pt]
%% Not sure if you wanted to add a sentence like this to note the non-polyconvexity of this model. I think it is important to acknowledge but note that we have the principal stretch model which still fits great if people insist on polyconvexity. 
%%%%%%%%%%%%%%%%%%%%%%%%%%%%%%%%%%%%%%%%%%%%%%%%%%%%%%%%%%%%%%%%%%%%%%%%%%%%
{\sffamily{\bfseries{Single-invariant-pricipal-stretch models perform well, but struggle in compression.}}}
%%%%%%%%%%%%%%%%%%%%%%%%%%%%%%%%%%%%%%%%%%%%%%%%%%%%%%%%%%%%%%%%%%%%%%%%%%%%
For the single-invariant-pricipal-stretch model 
with an $\textsf{L}_{0.5}$ regularization penalty of $\alpha = 1.0$, 
the discovered models for both foams 
fit the data accurately with only three terms. 
Their $\textsf{R}^2$ values 
range from 
$\textsf{R}^2=0.98$ to $\textsf{R}^2=1.00$ 
for both foams across all three experiments,
except for the compression fit of the \turbo
with $\textsf{R}_{\rm{com}}^2=0.94$.
For both foam models, 
a single principal-stretch term 
accounts for the majority of the stress 
in tension, compression, and shear. 
However, the additional single-invariant terms 
are necessary to achieve the appropriate shape 
of the compression curve. 
In particular, 
the stiffening of the foams at small stretches, 
which is potentially due to densification of the foam \cite{prabhakar_densification_2022}, 
cannot be modeled accurately
without these additional single-invariant terms. 
The transverse stress remains close to zero 
across the entire stretch regime, 
and does not exceed 15\,kPa in magnitude. 
%
%While it is appears that the transverse stress 
%may not remain zero outside the training regime, 
%this is not a major concern, 
%because we are already testing the foam 
%at much more extreme deformations 
%than those likely be observed in normal running. 
%
%Furthermore, we expect that a macroscopic material model that describes a cellular material may fail to describe the materials behavior at extreme deformations. 
\\[6.pt]
%%%%%%%%%%%%%%%%%%%%%%%%%%%%%%%%%%%%%%%%%%%%%%%%%%%%%%%%%%%%%%%%%%%%%%%%%%%%
{\sffamily{\bfseries{Limitations.}}}
%%%%%%%%%%%%%%%%%%%%%%%%%%%%%%%%%%%%%%%%%%%%%%%%%%%%%%%%%%%%%%%%%%%%%%%%%%%%
One limitation of all of our discovered models 
is that they are hyperelastic by design
and cannot predict the amount of energy 
dissipated during cyclic loading. 
While
both foams exhibit
low relative hysteresis,
9\% and 6\% in tension and 
10\% and 15\% in compression, 
we observe notable relative hysteresis of
26\% and 24\% in shear. 
To accurately model these effects, we need to  
collect additional data 
with varying loading rates and longer holding times 
\cite{boes_mechanics_2025},
and use inelastic constitutive neural networks \cite{holthusen_theory_2024}
to automatically discover 
inelastic models 
that explain these rate-dependent phenomena. 
Another limitation is that
our discovered models displays small transverse stresses,
which we could address with soft or hard constraints, 
for example, 
with input specific neural networks \cite{jadoon_isnn_2025}.
%%%%%%%%%%%%%%%%%%%%%%%%%%%%%%%%%%%%%%%%%%%%%%%%%%%%%%%%%%%%%%%%%%%%%%%%%%%%
\section{{\sffamily{\bfseries{Conclusion}}}}
\label{sec:conclusion}
%%%%%%%%%%%%%%%%%%%%%%%%%%%%%%%%%%%%%%%%%%%%%%%%%%%%%%%%%%%%%%%%%%%%%%%%%%%%
By performing a series of mechanical tests and 
using the data to train constitutive neural networks, 
we have discovered 
accurate, interpretable, and physics-informed 
constitutive models for 
two ultra-low density elastomeric polymeric foams,
the \leap and \turbo, 
used in the Asics Metaspeed \sky and \edge racing shoes. 
We tested five samples of each foam in 
uniaxial tension, 
unconfined and confined compression, 
and simple shear tests,
at a strain rate of 25\%/s, over 
a stretch range from 0.40 to 1.30 and 
a shear range from 0.00 to 0.25. 
The measurements revealed 
that both foams are highly compressible, 
exhibit pronounced tension–compression asymmetry, 
and possess an effective Poisson’s ratio
close to zero, 
consistent with their ultra-low density and high porosity.
By integrating the data into 
constitutive neural networks with sparse regression, 
we discovered compact, interpretable 
single-invariant models--supplemented 
by mixed-invariant or principal-stretch based terms--with 
$\textsf{R}^2$ values close to one across all loading modes. 
From a human performance perspective, 
these models enable finite element and gait-level simulations 
of carbon-fiber–plated racing shoes and
provide a quantitative framework 
for individualized assessment of running economy, 
performance outcomes, and injury risks
across varying running conditions.
From a basic-science perspective,
these results demonstrate 
that ultra-light cellular materials 
require constitutive descriptions 
that explicitly account for 
high compressibility and asymmetric deformation mechanisms 
beyond conventional single-invariant formulations.
More broadly, 
this work establishes a scalable and interpretable framework 
for constitutive modeling 
of highly compressible, ultra-light materials, 
including polymeric foams.
%%%%%%%%%%%%%%%%%%%%%%%%%%%%%%%%%%%%%%%%%%%%%%%%%%%%%%%%%%%%%%%%%%%%%%%%%%%%
\backmatter
%%%%%%%%%%%%%%%%%%%%%%%%%%%%%%%%%%%%%%%%%%%%%%%%%%%%%%%%%%%%%%%%%%%%%%%%%%%%
%%%%%%%%%%%%%%%%%%%%%%%%%%%%%%%%%%%%%%%%%%%%%%%%%%%%%%%%%%%%%%%%%%%%%%%%%%%%
\bmhead{Acknowledgements}
%%%%%%%%%%%%%%%%%%%%%%%%%%%%%%%%%%%%%%%%%%%%%%%%%%%%%%%%%%%%%%%%%%%%%%%%%%%%
This work was supported by the Wu Tsai Human Performance Alliance.
Jeremy McCulloch acknowledges 
the Wu Tsai Human Performance Alliance Digital Athlete Fellowship; 
Ellen Kuhl also acknowledges
the NSF CMMI grant 2320933 and 
the ERC Advanced Grant 101141626. \\[4.pt]
%%%%%%%%%%%%%%%%%%%%%%%%%%%%%%%%%%%%%%%%%%%%%%%%%%%%%%%%%%%%%%%%%%%%%%%%%%%%%
%%%%%%%%%%%%%%%%%%%%%%%%%%%%%%%%%%%%%%%%%%%%%%%%%%%%%%%%%%%%%%%%%%%%%%%%%%%%
%\bibliography{references}
%%%%%%%%%%%%%%%%%%%%%%%%%%%%%%%%%%%%%%%%%%%%%%%%%%%%%%%%%%%%%%%%%%%%%%%%%%%%

%%%%%%%%%%%%%%%%%%%%%%%%%%%%%%%%%%%%%%%%%%%%%%%%%%%%%%%%%%%%%%%%%%%%%%%%%%%%
\end{document}